\documentclass[11pt,a4paper,abstract=on]{scrartcl}

\usepackage[utf8]{inputenc}
\usepackage[english]{babel}

\usepackage[paper=a4paper,left=25mm,right=25mm,top=25mm,bottom=30mm]{geometry}
\usepackage{amsmath,amssymb,amsthm,mathrsfs,bbm,amsfonts,bm}
\usepackage{color,graphicx,cases,caption,enumerate}
\usepackage{float,multicol,authblk}
\usepackage[dvipsnames]{xcolor}
\usepackage{colortbl}
\usepackage{booktabs}
\usepackage{multirow}
\usepackage{setspace}
\usepackage{booktabs}
\usepackage{bbm}
\usepackage{capt-of}
\usepackage{array}
\usepackage[title]{appendix}
\newcolumntype{C}[1]{>{\centering\let\newline\\\arraybackslash\hspace{0pt}}m{#1}}
\usepackage{natbib}
\usepackage{placeins} 
\usepackage[breaklinks, colorlinks=true, citecolor=black, urlcolor=black, linkcolor=black]{hyperref}
\usepackage{subcaption}

\title{Uncertainty quantification for data-driven weather models}

\author[1]{Christopher Bülte\footnote{Current affiliation: Ludwig-Maximilians-Universität, Munich}}
\author[1]{Nina Horat}
\author[1]{Julian Quinting}
\author[1,2]{\\ Sebastian Lerch\footnote{corresponding author,  \texttt{sebastian.lerch@kit.edu}}}
\affil[1]{Karlsruhe Institute of Technology}
\affil[2]{Heidelberg Institute for Theoretical Studies}

\date{\today}

\begin{document}

\maketitle

\begin{abstract}
\noindent
Artificial intelligence (AI)-based data-driven weather forecasting models have experienced rapid progress over the last years. Recent studies, with models trained on reanalysis data, achieve impressive results and demonstrate substantial improvements over state-of-the-art physics-based numerical weather prediction models across a range of variables and evaluation metrics. Beyond improved predictions, the main advantages of data-driven weather models are their substantially lower computational costs and the faster generation of forecasts, once a model has been trained. However, most efforts in data-driven weather forecasting have been limited to deterministic, point-valued predictions, making it impossible to quantify forecast uncertainties, which is crucial in research and for optimal decision making in applications. Our overarching aim is to systematically study and compare uncertainty quantification methods to generate probabilistic weather forecasts from a state-of-the-art deterministic data-driven weather model, Pangu-Weather.
Specifically, we compare approaches for quantifying forecast uncertainty based on generating ensemble forecasts via perturbations to the initial conditions, with the use of statistical and machine learning methods for post-hoc uncertainty quantification. In a case study on medium-range forecasts of selected weather variables over Europe, the probabilistic forecasts obtained by using the Pangu-Weather model in concert with uncertainty quantification methods show promising results and provide improvements over ensemble forecasts from the physics-based ensemble weather model of the European Centre for Medium-Range Weather Forecasts for lead times of up to 5 days.
\end{abstract}

\section{Introduction}

Modern weather forecasts are usually based on simulations from physics-based numerical weather prediction (NWP) models, which describe atmospheric processes via systems of partial differential equations. 
To quantify forecast uncertainty and provide probabilistic predictions, NWP models are typically run several times with varying initial conditions and perturbed model physics, resulting in an ensemble of predictions. 
Numerically solving the differential equations requires tremendous computational resources, limiting the spatial resolution, as well as the number of ensemble runs. 
The history of NWP since its inception around 70 years ago has been a success story, albeit a ``quiet'' one characterized by continued, small improvements through the steady accumulation of scientific knowledge and technological advances \citep{Bauer2015}. 

Currently, a major leap in the formerly quiet success story of NWP can be observed due to the unprecedented success and rapid advancement of purely data-driven machine learning (ML) models for weather prediction. 
Contrary to NWP, data-driven weather models do not include any physics-based equations and aim to predict the future weather state (typically iteratively in steps of hours to days) from the initial weather state only, using statistical relations learned from past data. 
Beyond improved forecasts, the major advantages of data-driven models are their substantially lower computational costs (and accompanied energy consumption) and the faster generation of forecasts, once a model has been trained. 
Over the past two years, fundamental advances have been achieved, with purely data-driven weather models now convincingly outperforming state-of-the-art NWP systems, as recently reviewed in \citet{ECMWF2023rise}. 
The most notable contributions and global models include \citet{Keisler2022}, FourCastNet \citep{Pathak2022}, Pangu-Weather \citep{BiEtAl2023}, GraphCast \citep{LamEtAl2022}, ClimaX \citep{ClimaX}, FengWu \citep{FengWu}, FuXi \citep{FuXi}, SwinRDM \citep{SwinRDM}, AtmoRep \citep{AtmoRep}, NeuralGCM \citep{NeuralGCM}, Stormer \citep{Stormer}, GenCast \citep{GenCast}, and AIFS \citep{lang2024aifsecmwfsdatadriven}. 
All models utilize the ERA5 global reanalysis dataset \citep{era5} for training and evaluation, and are run at grid spacings of up to 0.25$^\circ$. 

However, most of these efforts have been focused on deterministic forecasts only, making it impossible to quantify forecast uncertainties which is crucial for optimal decision making, and one of the reasons underlying a transdisciplinary transition towards probabilistic forecasts \citep{GneitingKatzfuss2014}. 
Therefore, the overarching aim of our work is to investigate approaches to generate probabilistic predictions from deterministic data-driven weather models. 
An ideal solution to this challenge might be inherently probabilistic data-driven approaches, for example generative ML methods, and recent ensemble models such as AtmoRep \citep{AtmoRep}, NeuralGCM \citep{NeuralGCM}, GenCast \citep{GenCast} or FuXi-ENS \citep{FuXiENS} represent first steps in this direction. However, trained models are generally not yet publicly available and partly operate at different spatial resolutions than most of the deterministic data-driven weather models listed above. 

\begin{figure}[t]
	\centering
	\includegraphics[width = \textwidth]{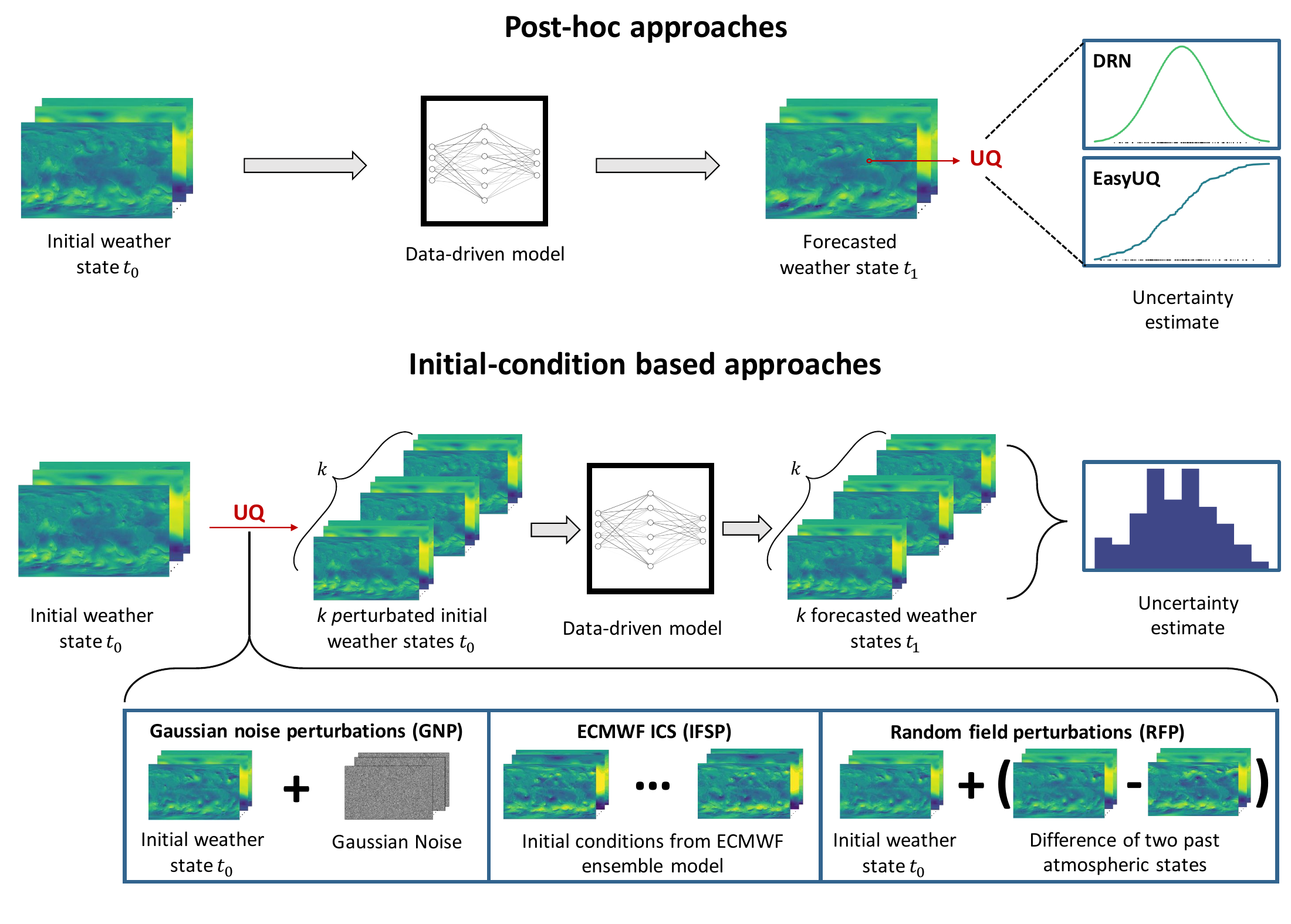}
	\caption{Schematic illustration of the different uncertainty quantification approaches to generate probabilistic forecasts from deterministic data-driven weather models. A detailed description of the UQ methods is provided in Section \ref{sec:methods}.}    \label{fig:overview}
\end{figure}

By contrast, we consider readily applicable techniques to generate probabilistic forecasts. Specifically, we consider two main approaches for uncertainty quantification (UQ) for data-driven weather models. A schematic overview of these approaches is provided in Figure \ref{fig:overview}.
\textit{Initial condition} (IC)-based approaches generate an ensemble forecast by running a data-driven model multiple times based on a number of (slightly) different initial conditions. 
These initial condition ensembles can be generated in various ways, and we consider three variants: Adding random noise to the initial weather state, as proposed for example by \cite{scher2021} or \cite{Pathak2022}; utilizing the perturbed initial conditions of a physics-based NWP ensemble model \citep{BuizzaEtAl2008}; and generating conditions based on perturbations computed from randomly selected past data, as proposed by \cite{MagnussonEtAl2009}.
IC approaches generally require the capabilities to run the data-driven models for a set of input data (i.e., estimated models, code, data, and suitable hardware infrastructure), which has recently become possible since code and data have been made public for some of the models (most notably FourCastNet, Pangu-Weather, and GraphCast), but still poses technical challenges due to the substantial computing and disk space requirements. 

\textit{Post-hoc} (PH) UQ approaches, by contrast, utilize statistical or ML methods to supplement deterministic forecasts with uncertainty information and thus turn them into probabilistic forecasts. These methods only require a training dataset of deterministic forecasts and corresponding observations.
A large variety of such approaches has been proposed, e.g., conformal prediction \citep{ConformalPrediction} or distributional regression \citep{GneitingKatzfuss2014}. 
Here, we consider two distributional regression approaches particularly relevant for atmospheric science applications, where such methods have been mostly used in the context of statistical post-processing to correct systematic errors of NWP forecasts \citep{VannitsemEtAl2021}.
The EasyUQ \citep{EasyUQ} approach builds on the recent isotonic distributional regression technique \citep[IDR;][]{IDR} and yields statistically optimal discrete predictive distributions by leveraging the pool-adjacent-violators algorithm for nonparametric isotonic regression.
EasyUQ utilizes deterministic forecasts of the target variable as sole input, and has, e.g., recently been used in \citet{EasyUQ_Africa} to generate probabilistic forecasts of precipitation from deterministic inputs. 
Over the past years, modern ensemble post-processing methods based on neural networks have been proposed which enable the incorporation of additional input variables and the data-driven learning of complex relationships between the inputs and distribution forecasts.
We will build on the parametric distributional regression network approach first proposed in \citet{RaspLerch2018}, which has been successfully extended for many target variables \citep[e.g., in][]{SchulzLerch2022} and has been used to generate corrected probabilistic forecasts from deterministic inputs from an NWP model \citep{ChapmanEtAl2022,GneitingEtAl2023}.

Our overarching aim is to systematically evaluate and compare the proposed UQ approaches for selected user-relevant target variables.\footnote{Note that recently, \citet{BrenowitzEtAl2024} proposed the use of a lagged ensemble of deterministic data-driven weather predictions as an alternative approach to obtain a probabilistic forecast. However, as argued by the authors themselves, this approach cannot be used for constructing real (out of sample) forecasts since it requires observations from a window around the initialization date, including initial conditions from the future. Additional adaptations of this approach thus seem necessary to enable a fair comparison to the UQ methods considered here.
}
We utilize the Pangu-Weather model \citep{BiEtAl2023} to produce deterministic and ensemble forecasts over Europe for a time period of five years, and conduct a systematic evaluation of the out-of-sample forecast performance of the various UQ approaches.
The operational ensemble forecast of the European Centre for Medium-Range Weather Forecasts (ECMWF) thereby serves as a benchmark model.

The remainder of this article is structured as follows. Section \ref{sec:setup} describes the data and setup of our case studies. Section \ref{sec:notation} introduces the notation and provides the mathematical formulation of the problem, and Section \ref{sec:methods} introduces the UQ methods, the predictive performance of which is evaluated in  Section \ref{sec:results}. Section \ref{sec:conclusion} concludes with a discussion.
Python code with implementations of all UQ methods is available at \url{https://github.com/cbuelt/dduq}.

\section{Data and setup}
\label{sec:setup}

Our study focuses on the Pangu-Weather model developed by \citet{BiEtAl2023}. 
Additional results for the FourCastNet model \citep{Pathak2022} are available in the supplemental material.
Pangu-Weather is a three-dimensional vision transformer architecture with specific adaptations and extensions to weather prediction, and was one of the first data-driven models to achieve improvements over physics-based NWP models.
The Pangu-Weather model produces global forecasts of five atmospheric variables (Z, Q, T, U, V) on 13 pressure levels and four surface variables (MSL, U10M, V10M, T2M) at a grid spacing of 0.25$^\circ$. 
It is trained based on 39 years of ERA5 reanalysis data from 1979--2017. In \citet{BiEtAl2023}, data from the year 2019 was used as validation data, and data from 2018 serves as a test dataset.
For details regarding the model architecture, training procedure, and forecast quality, we refer to \citet{BiEtAl2023}.
To implement the UQ methods described below, we adapted Pangu-Weather code and data provided by \citet{BiEtAl2023}\footnote{\url{https://github.com/198808xc/Pangu-Weather}} for our purposes.

Since some of the UQ methods discussed below require training and validation data on their own, we further produced both deterministic Pangu-Weather forecasts, as well as forecasts from the various UQ approaches, for additional recent years. 
In order to evaluate on data independent from the training data used in \citet{BiEtAl2023}, we utilize data from 2018--2021 as training and validation data for the UQ methods (if necessary), and evaluate all methods on data from 2022.

Forecasts from all methods are initialized at 00 UTC every day, for a total of $H = 31$ steps of 6 hours each (i.e., up to maximum lead time of 186 hours). 
Due to the substantial computing and disk space requirements (in particular when generating and storing ensemble forecasts), we restrict our attention to selected user-relevant weather variables (u-component and v-component of 10-m wind speed (U10 and V10), temperature at 2m and 850 hPa (T2M and T850), and geopotential height at 500 hPa (Z500)), and a European domain, covering an area from 35°N -- 75°N and 12.5°W -- 42.5°E. The ground truth for evaluation is the ERA5 dataset with a temporal resolution of 6 hours and a spatial grid spacing of 0.25°.

As a reference forecast but also as initial condition perturbations (see Section~\ref{sec:methods}), we retrieve operational ensemble forecasts of ECMWF's ensemble prediction system, which is based on the ECMWF Integrated Forecasting System (IFS). The data are retrieved for the same training and evaluation periods on a regular latitude-longitude grid of 0.25$\times$0.25° covering the identical spatial domain and forecast lead times. It should be noted that the native spatial resolution of the operational ensemble prediction system of ECMWF is slightly higher than that of ERA5 which may cause differences in regions of high topography. 
For comparisons with post-processed IFS predictions in Section \ref{sec:results}\ref{sec:IFSpp}, we further utilize IFS forecast data available in WeatherBench 2 \citep{WB2}.

\section{Mathematical notation}
\label{sec:notation}

In the following, we will consider probabilistic forecasts for several meteorological variables on a two-dimensional gridded domain. The grid point locations $(i,j), i = 1,...,I; j = 1,...,J$ will be summarized via a generic location index $l = 1,...,L$, where each value of $l$ denotes a specific combination of $i$ and $j$, and $L=IJ$. Where helpful, we will distinguish between a global domain $l\in\mathbb{L}_G$ and a European domain $l\in\mathbb{L}_E$, see Section \ref{sec:setup}.
The different target variables are treated separately and thus are omitted in the notation. 
Following common practice in NWP, we will consider forecasts to be initialized at time $t$, and to provide predictions for forecast horizons $h = 1,2,...,H$ steps ahead. 
In our case study, the forecast model runs will be started daily at 00 UTC and forecast steps will be 6 hours each. 
A deterministic Pangu-Weather (PW) forecast for location $l$, initialized at time $t$ and for a horizon of $h$ steps will be denoted by $X^\text{PW}_{l,t,h}$. 

Our overarching aim is to quantify forecast uncertainty in the form of a predictive distribution $F_{l,t,h}$. In most cases, this predictive distribution will be given in the form of a sample of size $M$, i.e., an ensemble forecast 
\[
\boldsymbol{X}_{l,t,h} = \left\{ X_{l,t,h}^1, \dots, X_{l,t,h}^M \right\},
\]
where, e.g., each ensemble member is started from a different set of initial conditions.

An observation corresponding to an $h$-step ahead forecast initialized at time $t$ is available at time $t+h$, and will be denoted by $Y_{l,t+h}$. 
As detailed below, we use the ERA5 reanalysis dataset as ground truth.
Since the deterministic data-driven model runs will typically be initialized from the corresponding ERA5 data at the initialization time, the starting conditions can be seen as 0-step ahead forecasts and will be denoted by $Y_{l,t}$.

\section{Methods}
\label{sec:methods}

This section provides a description of the different UQ methods we use to generate probabilistic forecasts from deterministic data-driven weather models. A schematic overview is available in Figure \ref{fig:overview}.

\subsection{Initial condition ensemble approaches}

The general idea behind all considered initial condition ensemble approaches is that based on (slightly) different initial conditions
\[
\boldsymbol{X}_{l,t,0} = \left\{ X_{l,t,0}^1, \dots, X_{l,t,0}^M \right\},
\]
an ensemble forecast of size $M$ is generated by starting $M$ runs of the deterministic data-driven weather model (Pangu-Weather in our case) from those initial conditions to produce an ensemble forecast
\[
\boldsymbol{X}_{l,t,h} = \left\{ X_{l,t,h}^1, \dots, X_{l,t,h}^M \right\} = \left\{ g_h(X_{l,t,0}^1), \dots, g_h(X_{l,t,0}^M) \right\}
\]
for $h=1,...,H$, where $g_h(X_{l,t,0}^m)$ denotes the $h$-step ahead Pangu-Weather forecast started from the IC ensemble member $ X_{l,t,0}^m$. 
Note that all IC approaches are based on generating global initial conditions for the full model grid and a global Pangu-Weather forecast is computed, even though we later restrict our attention to the grid over Europe for evaluation. The locations $l$ in the description of the IC approaches below should thus be understood as grid point locations of the global $0.25^\circ$ grid, $\mathbb{L}_G$.

\begin{figure}[h]
	\centering
	\begin{subfigure}[b]{\textwidth}
		\caption{GNP}
		\includegraphics[width=1\linewidth]{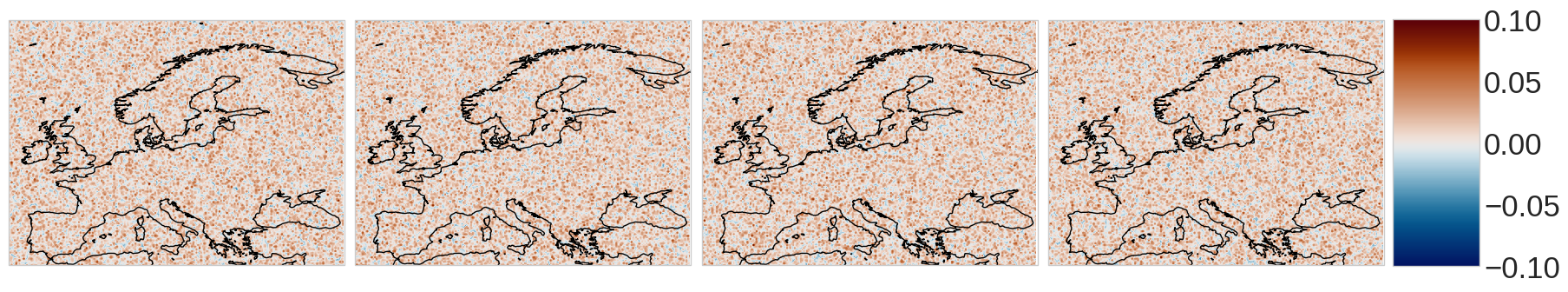}
	\end{subfigure}
	\begin{subfigure}[b]{\textwidth}
		\caption{RFP}
		\includegraphics[width=1\linewidth]{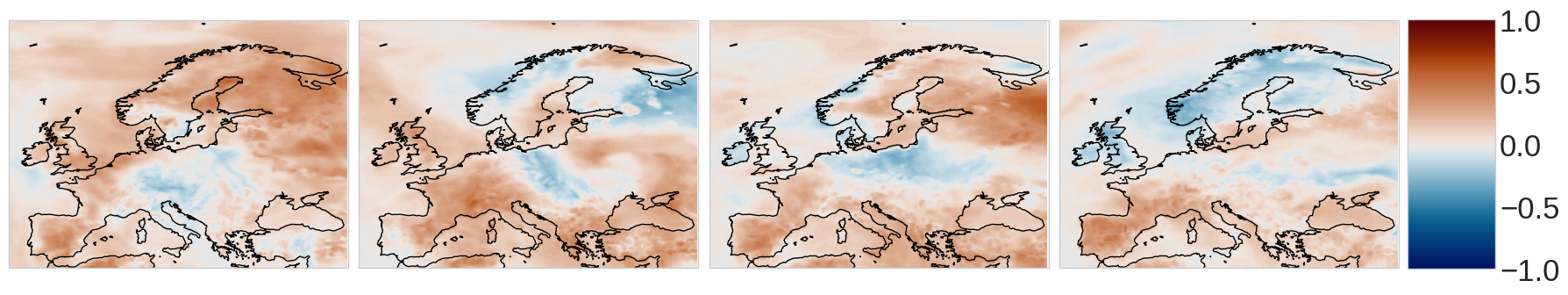}
	\end{subfigure}
	\begin{subfigure}[b]{\textwidth}
		\caption{IFSP}
		\includegraphics[width=1\linewidth]{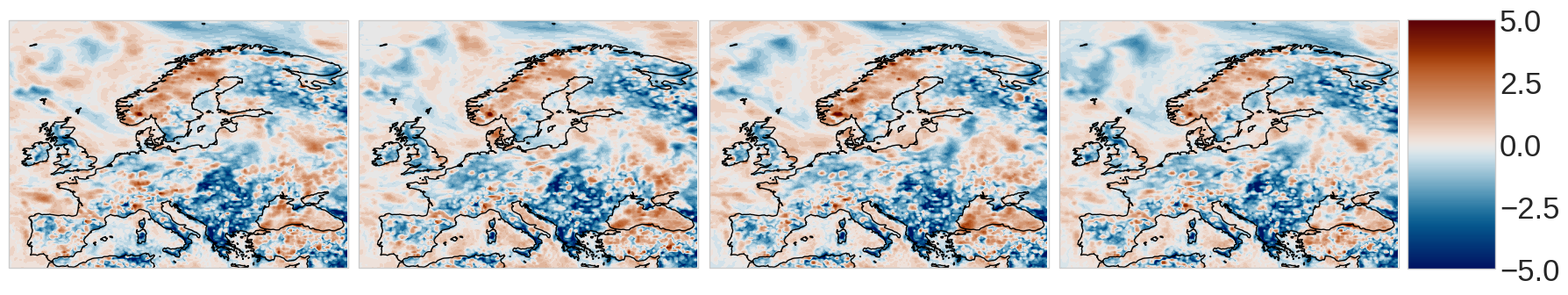}
	\end{subfigure}
	\caption{Exemplary perturbations of the different initial condition ensemble approaches across the European domain. Each row shows the residual to the original ERA5 observation for forecasts initialized on June 1, 2022, the variable T2M, and four randomly selected perturbations.}
	\label{fig:perturbation_example}
\end{figure}

The IC approaches described below mainly differ in the way the IC ensemble $\boldsymbol{X}_{l,t,0}$ is generated. Specifically, we consider Gaussian noise perturbations, random field perturbations and IFS perturbed initial conditions, which are introduced in detail below. 
Exemplary perturbations for a common initialization date are visualized in Figure \ref{fig:perturbation_example}. 
A natural shortcoming of all IC approaches is that they are inherently limited to accounting for initial condition uncertainty only, and not for model uncertainty, which is, e.g., addressed in physics-based NWP models via stochastic parametrizations of subgrid processes \citep{Palmer2019param}. One approach to address this has recently been proposed by \citet{MaheshEtAl2024}, who construct IC perturbations with bred vectors and incorporate model uncertainty by utilizing an ensemble of data-driven weather models, the members of which have been trained separately from different random starting points.

\subsubsection{Gaussian noise perturbations (GNP)}

A simple and straightforward method to generate an IC ensemble is to add random noise to the ERA5-based initial weather state $Y_{l,t}$ from which the deterministic Pangu-Weather model would be initialized. We here follow  \citet{Pathak2022}, who first proposed this approach for the FourCastNet model, and generate inital conditions by adding independently sampled Gaussian noise to all variables after standardization, i.e.,
\[
X_{l,t,0}^{\text{GNP},m} = Y_{l,t} + \varepsilon_{l,t}^m \quad \text{for } m = 1,...,M,
\]
where $\varepsilon_{l,t}^m \sim \mathcal{N}(0,\gamma\,\sigma)$, $\sigma$ denotes the mean standard deviation of the respective variable over all grid points, and $\gamma$ is a tuning parameter. Samples of the Gaussian noise process are thus generated independently over members $m$, locations $l$, variables and initialization times $t$. While \citet{Pathak2022} use $\gamma = 0.3$ for the FourCastNet model, the Pangu-Weather model utilizes a substantially different architecture and models an increased number of meteorological variables. For our experiments we found that a scaling factor value of $\gamma = 0.001$ applied to all variables works sufficiently well.

Alternative specifications of the noise process have been considered. For example, \citet{BiEtAl2023} use Perlin noise, but our initial experiments indicated only negligible differences to the performance of Gaussian noise-based GNP forecasts for our case study. \citet{GenCast} recently proposed a noise process where spatial dependencies on the sphere are retained in the context of a generative data-driven weather model, which might constitute an interesting alternative. 

\subsubsection{IFS initial conditions (IFSP)}\label{sec:ifsp}

Further, we consider an initial condition approach more akin to the operational practice of running NWP ensemble models \citep{palmer2019} by utilizing the (ensemble of) initial conditions of the ECMWF ensemble prediction system to initialize the deterministic data-driven model.
Specifically, we select the values at initialization time (i.e., the forecasts at step $h=0$) of the perturbed members $\boldsymbol{Z}_{t,0}^m = \{ Z_{l,t,0}^m, l \in \mathbb{L}_G \}, m = 1,...,M$, of the ECMWF ensemble, i.e.,
\[
X_{l,t,0}^{\text{IFSP},m} = Z_{l,t,0}^m \quad \text{for } m = 1,...,M. 
\]
For the period 2018--2022, 
we remapped the initial conditions from a native grid spacing 
to a regular latitude-longitude grid of 0.25° grid spacing.

The initial condition uncertainty in ECMWF's ensemble prediction system is incorporated by two approaches. 
The ensemble of 4D-var data assimilations generates 25 independent ensemble members by introducing perturbations to observations, physical processes in the short-term forecasts and the sea surface temperature state \citep{Isaksen2010}. Further, singular vector perturbations are added to the analysis field which lead to a rapid dispersion of the ensemble members \citep{Leutbecher2008}. Accordingly, one would expect faster dispersion of ensemble members than with Gaussian perturbations. 

\subsubsection{Random field perturbations (RFP)}

Finally, an alternative data-driven approach to generate IC ensembles, which we will refer to as random field perturbations, was proposed by \citet{MagnussonEtAl2009} in the context of physics-based NWP ensemble models. They argue that adding noise to the initial conditions ignores the underlying dynamics of the weather system. Instead, they suggest to use the scaled difference of two independent, randomly selected atmospheric states from the past as perturbation, which has the advantage of preserving linear balances in the system. 
The random field perturbations are calculated as
\begin{equation*}
\boldsymbol{\xi}_t^{m,\alpha} = \alpha \frac{\boldsymbol{Y}_{\tau_1^m}- \boldsymbol{Y}_{\tau_2^m}}{\|\boldsymbol{Y}_{\tau_1^m}- \boldsymbol{Y}_{\tau_2^m} \|_{\mathrm{Etot}}} \quad \text{for } m = 1,...,M,
\end{equation*}
where $\boldsymbol{Y}_{\tau_i^m} = \{ Y_{l,\tau_i^m}, l\in \mathbb{L}_G\}, i = 1,2$ denotes the global observed ERA5 field of the selected variables at date $\tau_i^m$, and $\|\cdot\|_{\mathrm{Etot}}$ denotes the total energy norm, which is a conserved quantity of the governing equations of motion linearized about a reference state \cite[cf.][]{MagnussonEtAl2009}. 
We choose the dates $\tau_1^m, \tau_2^m$ randomly from the training dataset (2018--2021) and from the same month as $t$ to account for seasonal variability, but sample $\tau_1^m$ and $\tau_2^m$ from different years to ensure (approximate) independence.
The constant $\alpha$ is a tuning parameter and controls the dispersion of the IC ensemble. 
Based on preliminary tests for a subset of initilization dates in which we tested the sensitivity of the spread-skill relationship to the magnitude of $\alpha$, we chose $\alpha = 5\cdot 10^{6}$ for our case study. Increasing or decreasing the magnitude of $\alpha$ deteriorated the spread-skill relationship. Note that despite this scaling, the initial perturbation in terms of total energy are greater than with IFS initial conditions \citep{MagnussonEtAl2009}.
With these choices, global perturbations $\boldsymbol{\xi}_{t}^{m,\alpha}, m = 1,...,M$, are computed and added to the corresponding ERA5 initial conditions, i.e.,
\[
X_{l,t,0}^{\text{RFP},m} = Y_{l,t} + \xi^{m,\alpha}_{l,t}
\]
for $m = 1,...,M$ and all $l \in \mathbb{L}_G$.

\subsection{Post-hoc approaches}

In contrast to the IC approaches, the PH methods operate directly on a given deterministic forecast from a data-driven weather model, and learn from past pairs of forecasts and observations how to best generate a probabilistic forecast from the deterministic input. 
From a meteorological perspective, this can be viewed as a post-processing task \citep{VannitsemEtAl2021}.
In the following, we assume that a dataset of past deterministic Pangu-Weather forecasts and corresponding observations, 
\[
\left(X^\text{PW}_{l,t,h}, Y_{l,t+h}\right), \quad \text{for } l\in\mathbb{L}_E,
\]
is available, where $t$ denotes an initialization time in the training dataset (2018--2021).

Based on the training dataset, the PH methods yield forecast distributions $F_{l,t,h}$. 
Given a deterministic Pangu-Weather forecast $X^\text{PW}_{l,t^\ast,h}$ in the test dataset (2022), a probabilistic forecast for the date $t^\ast$ and lead time $h$ can thus be obtained by using the corresponding deterministic forecast as input to the trained PH model.
In the following, we consider two complementary post-hoc methods based on statistical and ML approaches.

An advantage of the PH methods compared to the IC approaches is their ability to correct systematic errors such as biases in the deterministic forecasts, and that they are not limited to accounting for initial condition uncertainty only. 
However, these methods require sufficient training data to generate forecasts, unlike, e.g., the GNP and IFSP approaches. 
We here utilize four years of training data to ensure a separation to the data used to train the Pangu-Weather model by \citet{BiEtAl2023}. In principle, larger training datasets could be obtained by generating Pangu-Weather forecasts for the preceding years, at the potential risk of overfitting.

\subsubsection{EasyUQ}

EasyUQ, proposed by \cite{EasyUQ}, aims at learning a predictive distribution from deterministic, single-valued model output.
As noted in the introduction, EasyUQ is a special case of IDR \citep{IDR} for a single deterministic prediction.
EasyUQ proceeds separately for every location $l\in\mathbb{L}_E$ and lead time $h$. 
To simplify notation, we will suppress the lead time index $h$ in the current subsection, and note that all forecasts and observations should be understood as those for the corresponding lead time only.
Given corresponding data of the form $(X^\text{PW}_{l,t}, Y_{l,t}), t = 1, ..., T$, and assuming that the predictive cumulative distribution functions (CDFs) $F_{x}(y) = \mathbb{P}(Y_{l,t} \leq y | X_{l,t} = x)$ are increasing in stochastic order in $x$, i.e., $F_x(y) \geq F_{x^\prime}(y)$ for all $y\in\mathbb{R}$ if $x\leq x^\prime$, the EasyUQ-estimated predictive CDF is then given by 
\[
\hat{F}^\text{EasyUQ}_{l,t} (y) := \hat{F}^\text{EasyUQ}_{X^\text{PW}_{l,t}} (y) = \min_{k=1,\ldots, t} \max_{\ell=t,\ldots,T} \frac{1}{\ell-k+1} \sum_{t^\prime=k}^\ell \mathbb{I} \{Y_{l,t^\prime} \leq y \}, \quad t = 1,\ldots, T.
\]

Thereby, $\hat{F}^\text{EasyUQ}_{l,t}$ is a statistically optimal discrete predictive distribution in that it minimizes the continuous ranked probability score (CRPS, see Section \ref{sec:methods}\ref{sec:eval}) over all conditional distributions satisfying the assumption of stochastic ordering. 
For theoretical results and more details on EasyUQ, we refer to \cite{EasyUQ} and \citet{IDR}.
EasyUQ does not require any choices of tuning parameters and thus constitutes an attractive benchmark method that can be applied in a fully automated manner.
\citet{EasyUQ} note that EasyUQ yields similar forecast performance as conformal prediction, and in case studies on post-processing, EasyUQ showed predictive performance comparable to other statistical methods \citep[e.g.,][]{SchulzLerch2022}.

\subsubsection{Distributional regression network (DRN)}

A key limitation of the EasyUQ approach is that there is no straightforward way to include additional predictor variables besides the deterministic forecasts of the target variable of interest. 
However, recent research on ML-based ensemble post-processing methods has highlighted that incorporating additional predictors is a key aspect in the substantial improvements achieved by these approaches \citep{RaspLerch2018}. 
Therefore, we consider a DRN, a parametric neural network (NN)-based approach first proposed in \citet{RaspLerch2018} as an alternative PH method. 
To introduce DRN, we slightly extend the notation from above and use $\mathbb{X}_{l,t,h}^\text{PW}$ to denote the (vector of) deterministic Pangu-Weather forecasts for all considered output variables at location $l\in\mathbb{L}_E$, initialization time $t$ and lead time $h$. Note that we consider only predictor variables in the DRN from the same vertical atmospheric level as the target variable of interest. Extensions towards incorporating additional predictors from other vertical levels is left for future work.  

Based on the deterministic model output, $\mathbb{X}_{l,t,h}^\text{PW}$, the DRN approach proceeds by training a NN which yields the parameters of a suitable parametric distribution for the target variable as its output.
DRN enables the use of arbitrary predictors as inputs to the NN, including additional meteorological variables (in our case from Pangu-Weather outputs) and location information (i.e., latitude and longitude). 
The NN parameters are determined by minimizing the CRPS over the training dataset.
As for EasyUQ, we estimate separate models for every lead time, but note that considering multiple lead times jointly can be a viable alternative \citep{PrimoEtAl2024}.
In our case study, we use a Gaussian predictive distribution for all target variables and closely follow \citet{RaspLerch2018} in our implementation.
We fit a single DRN model for each forecast horizon $h$ jointly over all grid points $l\in\mathbb{L}_E$ in the target domain.
In addition to the deterministic Pangu-Weather forecasts $\mathbb{X}_{l,t,h}^\text{PW}$, the locations are encoded via a positional embedding that maps a location $l\in\mathbb{L}_E$ to a vector of latent features, which are then used as auxiliary input variables of the NN. This procedure aims at making the model locally adaptive, while avoiding the training of a separate model at every grid point. 
The DRN model thus yields a predictive distribution 
\[
F^\text{DRN}_{l,t,h} = \mathcal{N}_{\mu_{l,t,h}, \sigma_{l,t,h}}
\]
for each location  $l\in\mathbb{L}_E$ and lead time $h$, where $\mu_{l,t,h}$ and $\sigma_{l,t,h}$ are the location and scale parameter of the Gaussian forecast distribution obtained as output of the NN.

We fit separate DRN models for all target variables, and use an identical NN architecture, the hyperparameters of which were determined based on a limited series of initial tuning experiments.
Specifically, we use a NN with a single hidden layer of size 512, a location embedding of dimension 5, a batch size of 1024, and train the model for 30 epochs.
In principle, it might be possible to improve the predictive performance of the DRN models further, e.g., by a more extensive hyperparameter search. 
However, the forecast performance of DRN models has been demonstrated to be fairly robust in this regard \citep[e.g.,][]{SchulzLerch2022}.

\subsection{Forecast evaluation methods} \label{sec:eval}

To compare the various UQ methods introduced above, we mainly rely on comparing their out of sample forecast performance based on proper scoring rules \citep{gneiting2007}.
Proper scoring rules enable a simultaneous assessment of calibration and sharpness of a probabilistic forecast, and have become widely used across disciplines. 
Generally, a scoring rule $S$ assigns a numerical score to a predictive distribution $F$ and a corresponding realized observation $y$, and is called proper, if in expectation, the true distribution of the observation receives the best possible (i.e., minimal) score, i.e.,
\[
\mathbb{E}_{Y\sim G} S(G,Y) \leq \mathbb{E}_{Y\sim G} S(F,Y) \quad \text{for all } F,G\in\mathcal{F},
\]
where $\mathcal{F}$ denotes a suitable class of forecast distributions, see \citet{gneiting2007} for details.
A commonly used scoring rule for evaluating univariate probabilistic forecasts in meteorological applications is the continuous ranked probability score \citep[CRPS,][]{MathesonWinkler1976}, 
\begin{equation*}
\label{eq:crps}
\mathrm{CRPS}(F, y) = \int_{- \infty}^\infty \big ( F(z) - \mathbb{I} \{ y \leq z \} \big )^2 dz,
\end{equation*}
where $\mathbb{I}$ denotes the indicator function, and $F$ is assumed to have a finite first moment. 
Closed-form expressions of the CRPS are available for both sample-based predictive distributions in the form of an $M$-member ensemble, as well as for many parametric families of forecast distributions, see, e.g., \citet{scoringRules}.

Skill scores based on proper scoring rules are a common tool to assess the relative improvements over a reference forecasting method. The continuous ranked probability skill score (CRPSS) is given by 
\begin{equation*}
\mathrm{CRPSS}_F = \frac{\mathrm{\overline{\mathrm{CRPS}}_{\mathrm{ref}}}-\overline{\mathrm{CRPS}}_F}{\mathrm{\overline{\mathrm{CRPS}}_{\mathrm{ref}}}},
\end{equation*}
where $\overline{\mathrm{CRPS}}_F$ denotes the average CRPS of $F$ over a test set, and $\overline{\mathrm{CRPS}}_{\mathrm{ref}}$ denotes the corresponding average CRPS of a reference method. 
The CRPSS is positively oriented, with negative values indicating worse performance than the reference, 0 indicating no improvement, and a maximum value of 1.

Further, we employ probability integral transform \citep[PIT,][]{gneiting.2007b} histograms to assess the calibration of the probabilistic forecasts. The PIT $F(y)$ is the value the predictive CDF $F$ of the forecast obtains at the realized outcome $y$. 
For a calibrated forecast, the PIT should follow a uniform distribution and corresponding deviations can be attributed to specific types of miscalibration \citep{gneiting.2007b}.
In addition, we further assess the reliability of the ensemble forecasts by comparing the root mean squared error (RMSE) of the ensemble mean to the average ensemble spread, which should approximately be equal across time for a calibrated ensemble \citep{FortinEtAl2014}.

\section{Results}
\label{sec:results}

\subsection{UQ methods applied to Pangu-Weather}

\begin{table}
	\centering
	\caption{Mean CRPS of all methods and variables across the European domain for three different groups of lead times, with the best-performing method highlighted in bold. Note that the CRPS values for Z500 are scaled by a factor of $0.01$.}
	\label{tab:results_pangu}
	\begin{tabular}{ c c|c| c c c c c } 
		\toprule
		& Variable & ECMWF IFS  & GNP & IFSP &  RFP & EasyUQ & DRN \\
		\midrule
		\multirow{5}{5em}{6h - 48h} & U10 & 0.54 & 0.71 & 0.78 & 0.58 & 0.53 & \textbf{0.51} \\
		&V10  & 0.54 & 0.71 & 0.79 & 0.58 & 0.53 & \textbf{0.51} \\
		&T2M  & 0.57 & 0.60 & 0.83 & 0.50 & 0.43 & \textbf{0.41} \\
		&T850 & 0.43 & 0.57 & 0.81 & 0.45 & 0.43 & \textbf{0.41} \\
		&Z500 & 0.33 & 0.48 & 1.71 & 0.36 & 0.36 & \textbf{0.32} \\
		\midrule
		
		\multirow{5}{5em}{48h - 120h} & U10 & \textbf{0.96} & 1.39 & 1.54 & 1.03 & 1.05 & 1.03 \\
		& V10 & \textbf{0.96} & 1.39 & 1.57 & 1.02 & 1.05 & 1.03 \\
		& T2M & 0.75 & 0.96 & 1.22 & 0.74 & 0.69 & \textbf{0.67} \\
		& T850 & \textbf{0.75} & 1.09 & 1.38 & 0.80 & 0.82 & 0.79 \\
		& Z500 & \textbf{1.21} & 1.75 & 2.52 & 1.26 & 1.35 & 1.29 \\
		\midrule
		
		\multirow{5}{5em}{$\geq$ 120h} & U10 & \textbf{1.54} & 2.31 & 2.09 & 1.59 & 1.70 & 1.68 \\
		& V10 & \textbf{1.58}& 2.36 & 2.17 & 1.62 & 1.74 & 1.71 \\
		& T2M & \textbf{1.05}& 1.51 & 1.57 & 1.07 & 1.13 & 1.10 \\
		& T850 &\textbf{1.33} & 2.01 & 2.04 & 1.39 & 1.55 & 1.48 \\
		& Z500 & \textbf{2.91} & 4.29 & 4.73 & 3.00 & 3.36 & 3.25 \\
		\bottomrule
	\end{tabular}
\end{table}

We here compare the previously introduced UQ methods based on their out-of-sample predictive performance. 
The ensemble forecasts from the operational 50-member ECMWF model are used as a baseline, and can be considered as a state-of-the-art physics-based NWP ensemble model.
Note that we did not apply any post-processing to the ECMWF ensemble forecasts here, but compare selected UQ methods to post-processed deterministic IFS forecasts in the next subsection. 
The evaluation and training setup follows the descriptions in Section \ref{sec:setup}, and we generate ensembles of size $M = 50$ for all initial condition ensemble approaches. The PH methods are evaluated based on their predictive distributions, i.e., the empirical CDF for EasyUQ and the Gaussian forecast distribution for DRN. 

\begin{figure}[h]
    \centering
    \includegraphics[width = \textwidth]{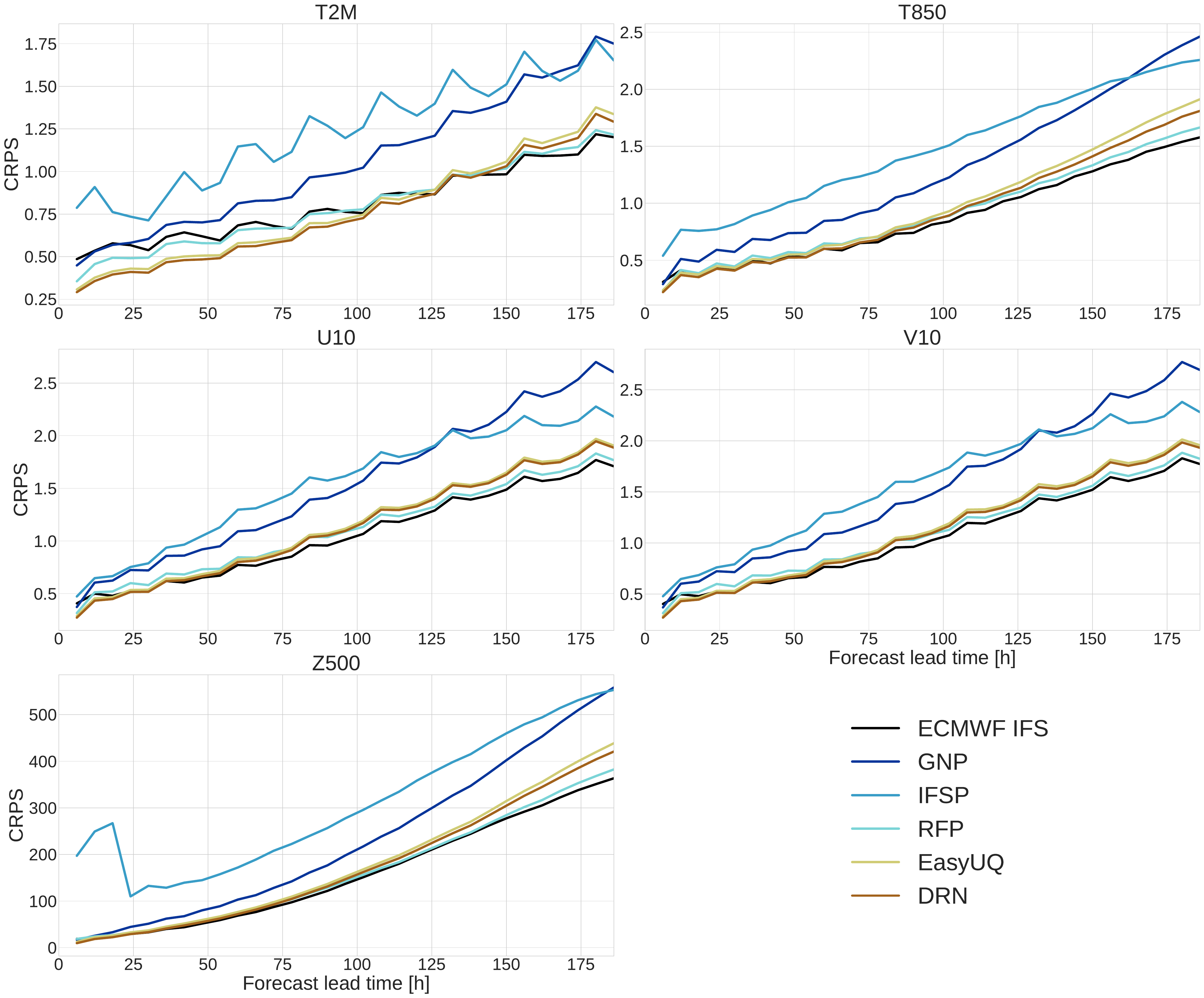}
    \caption{Mean CRPS as a function of the forecast lead time for the different UQ methods, aggregated over all locations.}
    \label{fig:crps_per_time_pangu}
\end{figure}

Table \ref{tab:results_pangu} provides the mean CRPS for all UQ methods, averaged over all grid points in the European domain and stratified into three groups of forecast lead times. 
Figure \ref{fig:crps_per_time_pangu} shows the mean CRPS as a function of the forecast lead time for all variables.
Both illustrations indicate that the use of data-driven weather forecasts in concert with the PH methods proposed here can yield improvements over the ECMWF ensemble forecasts.
The extent of these improvements, and the relative performance of the different UQ methods, strongly depends on the variable of interest as well as the forecast lead time.
Generally, DRN yields the best forecasts at shorter lead times, followed closely by the EasyUQ model.
For longer lead times up to 120\,h, the CRPS of the ECMWF ensemble is similar to that of the DRN, EasyUQ and RFP approaches, and for lead times beyond 120\,h, the ECMWF ensemble performs better than all compared UQ methods.
The most pronounced differences and most clear improvements over the ECMWF ensemble forecasts can be observed for T2M.
The rankings among the different UQ methods are mostly identical across the considered target variables, with the PH methods (DRN and EasyUQ) showing better forecasts at shorter lead times, whereas the RFP approach yields the best forecasts at longer lead times.  
Interestingly and in contrast to previous studies on ensemble post-processing \citep[e.g.,][]{SchulzLerch2022}, DRN only yields relatively minor improvements over the considerably simpler EasyUQ method. A potential explanation might be that DRN here is restricted to much fewer additional predictor variables compared to other studies. Therefore, further improvements might be achieved by, e.g., incorporating Pangu-Weather outputs from vertical atmospheric levels other than the vertical level of the target variable itself.
The GNP and IFSP approaches lead to substantially worse forecasts compared to the other methods for all variables, in particular for Z500.

\begin{figure}[p]
\centering
\begin{subfigure}[b]{\textwidth}
    \caption{U10}
   \includegraphics[width=1\linewidth]{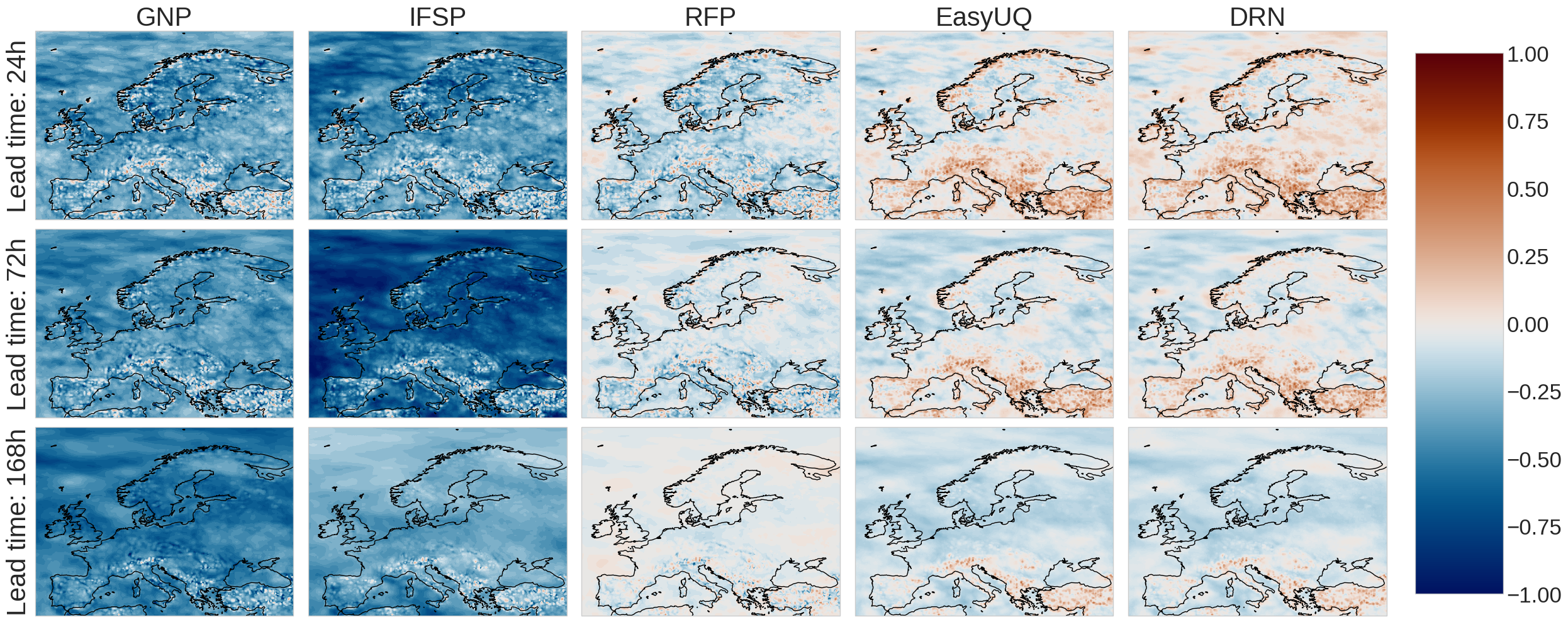}
\end{subfigure}
\begin{subfigure}[b]{\textwidth}
    \caption{T2M}
   \includegraphics[width=1\linewidth]{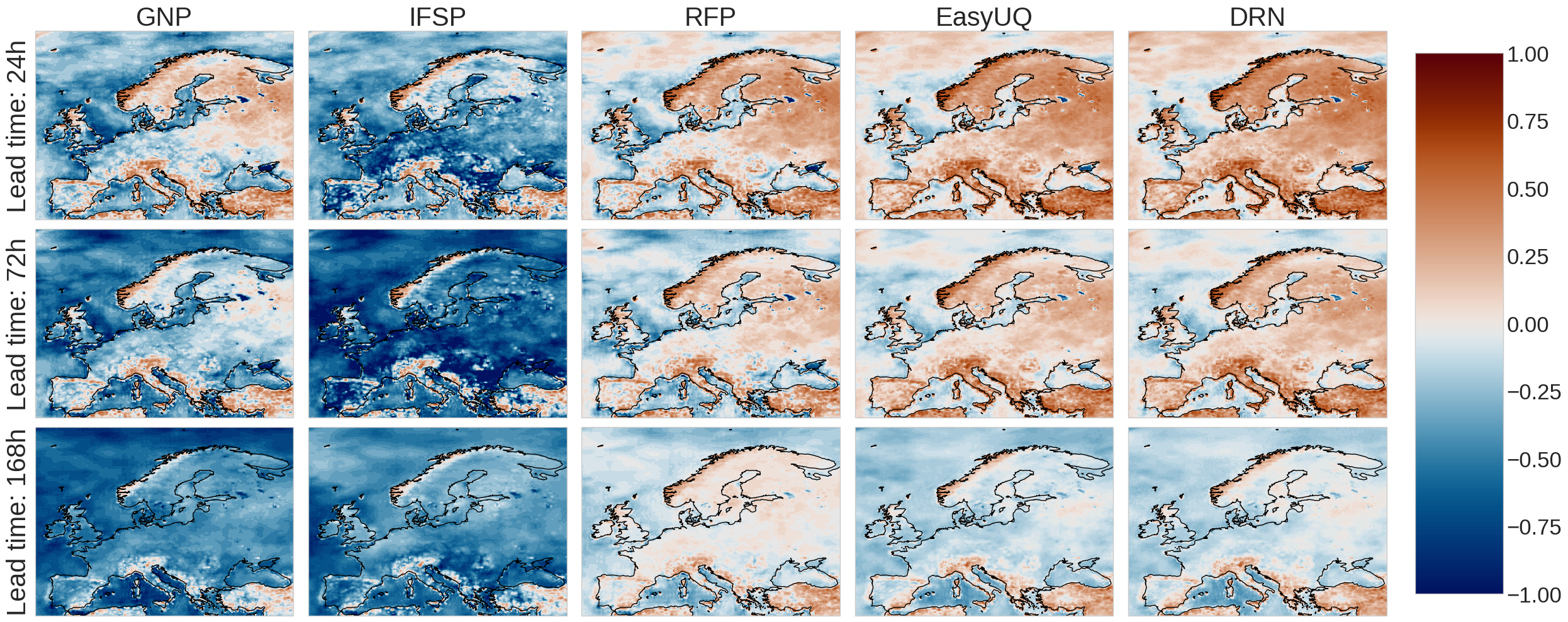}
\end{subfigure}
\begin{subfigure}[b]{\textwidth}
    \caption{Z500}
   \includegraphics[width=1\linewidth]{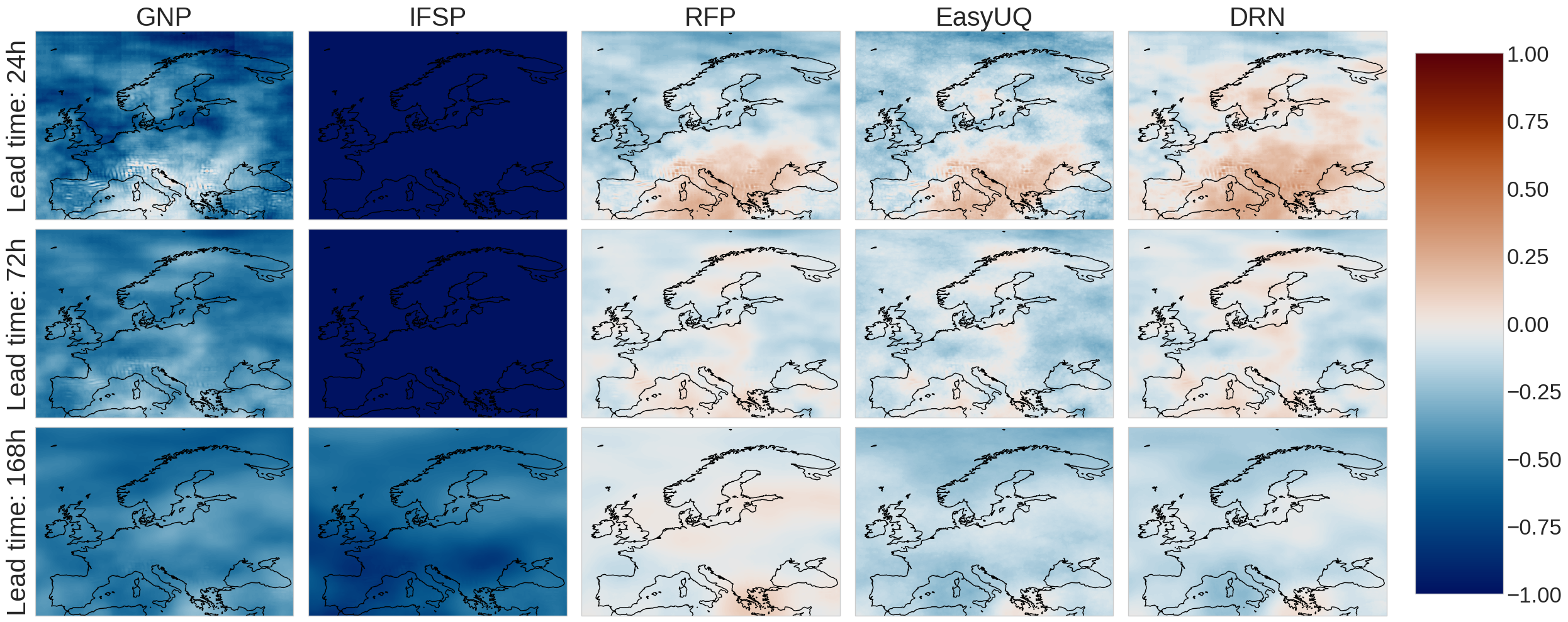}
\end{subfigure}
\caption{CRPSS of the different UQ methods over the spatial domain, using the ECMWF ensemble as a reference method. The rows correspond to specific forecasting lead times. Note that positive CRPSS values indicate an improvement over the reference in terms of the CRPS at the respective grid point.}
\label{fig:pangu_crpss}
\end{figure}

Figure \ref{fig:pangu_crpss} shows the CRPSS of the different UQ methods over the spatial domain, using the ECMWF ensemble as a reference forecast.
For most methods, target variables and lead times, there are some geographical regions where improvements over the ECMWF ensemble are obtained.
The most notable improvements and variations can be observed for the variable T2M. 
There, the improvements over the ECMWF ensemble forecasts are most pronounced over land, and even for a lead time of 168\,h, all methods show a positive skill score over mountainous regions.
In particular, the PH methods indicate a notably better performance. 
A generally similar spatial pattern, albeit with less pronounced improvements over the ECMWF ensemble, can be observed for U10. 
The areas with positive CRPSS values for Z500 seem to be more concentrated around the Mediterranean and south-eastern Europe, and the GNP and IFSP methods perform notably worse.

\begin{figure}[p]
\centering
\begin{subfigure}[b]{\textwidth}
    \caption{U10}
   \includegraphics[width=1\linewidth]{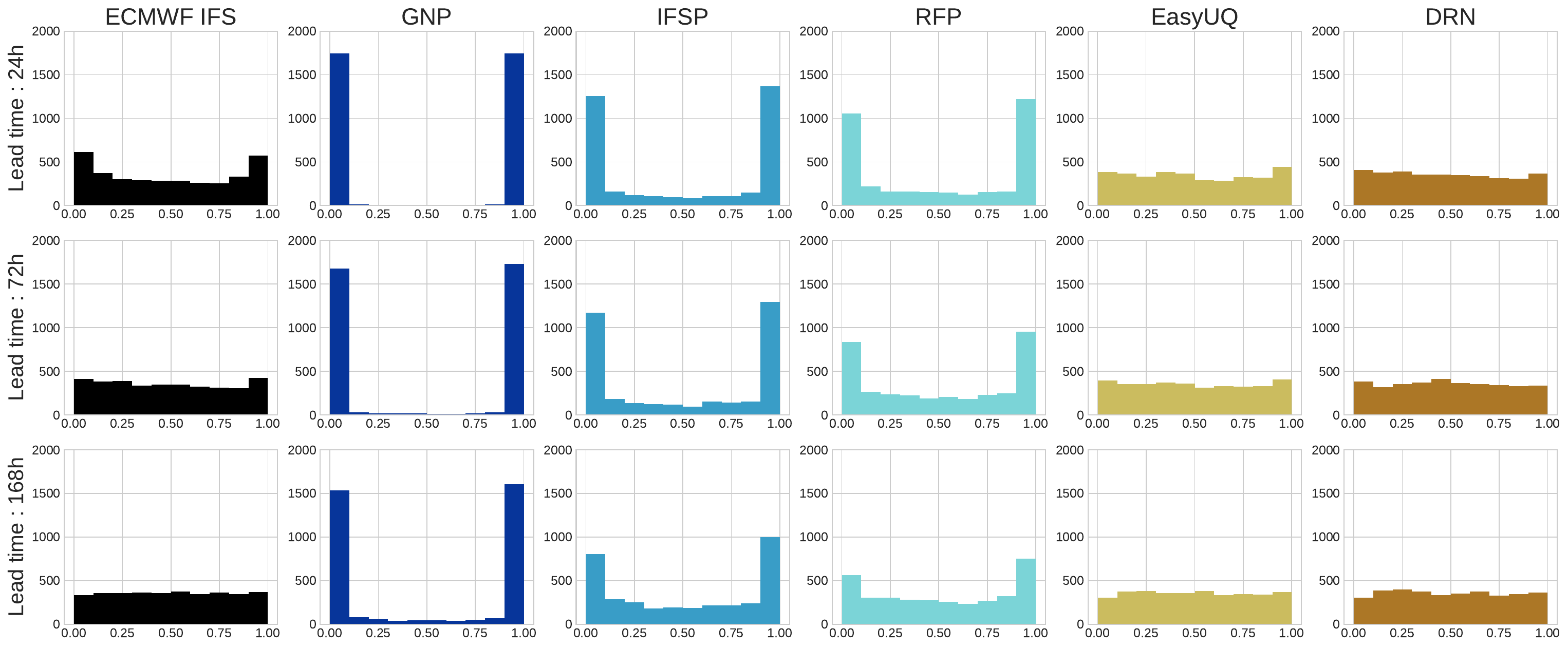}
\end{subfigure}
\begin{subfigure}[b]{\textwidth}
    \caption{T2M}
   \includegraphics[width=1\linewidth]{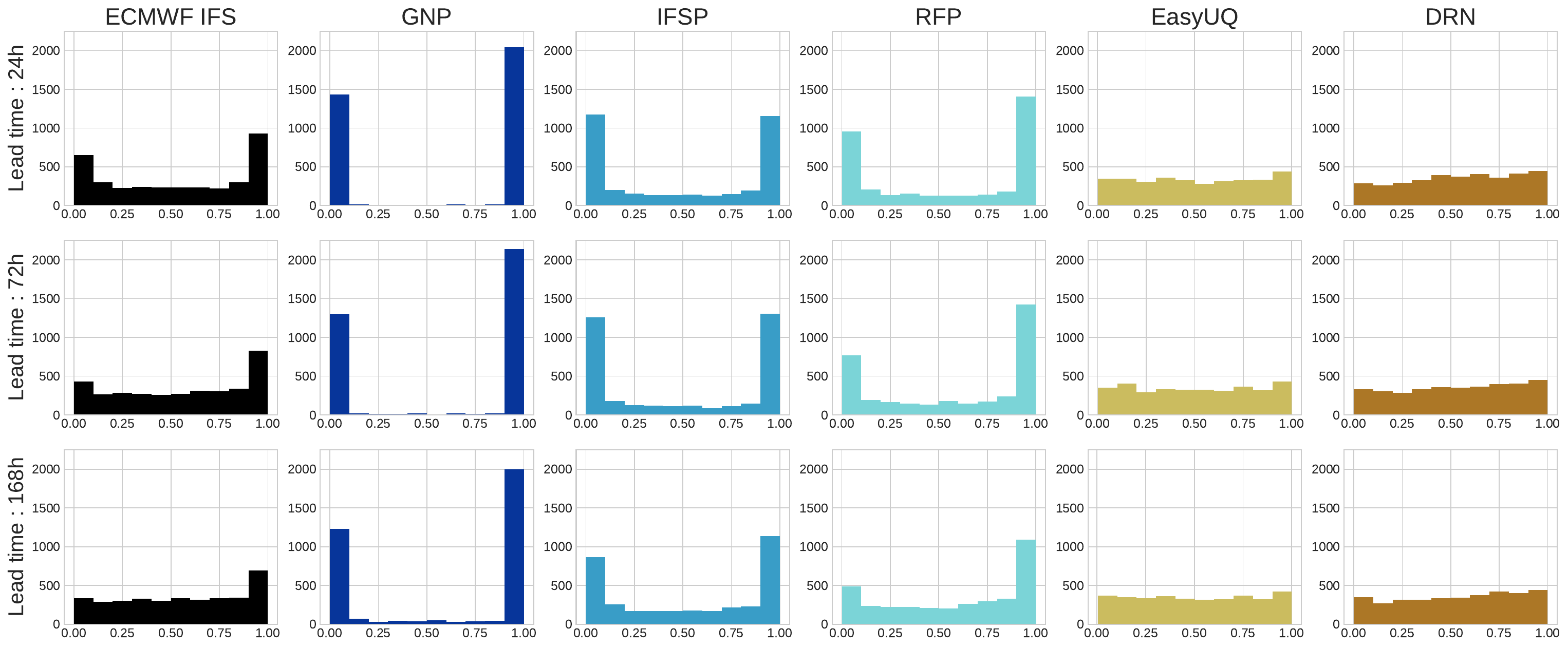}
\end{subfigure}
\begin{subfigure}[b]{\textwidth}
    \caption{Z500}
   \includegraphics[width=1\linewidth]{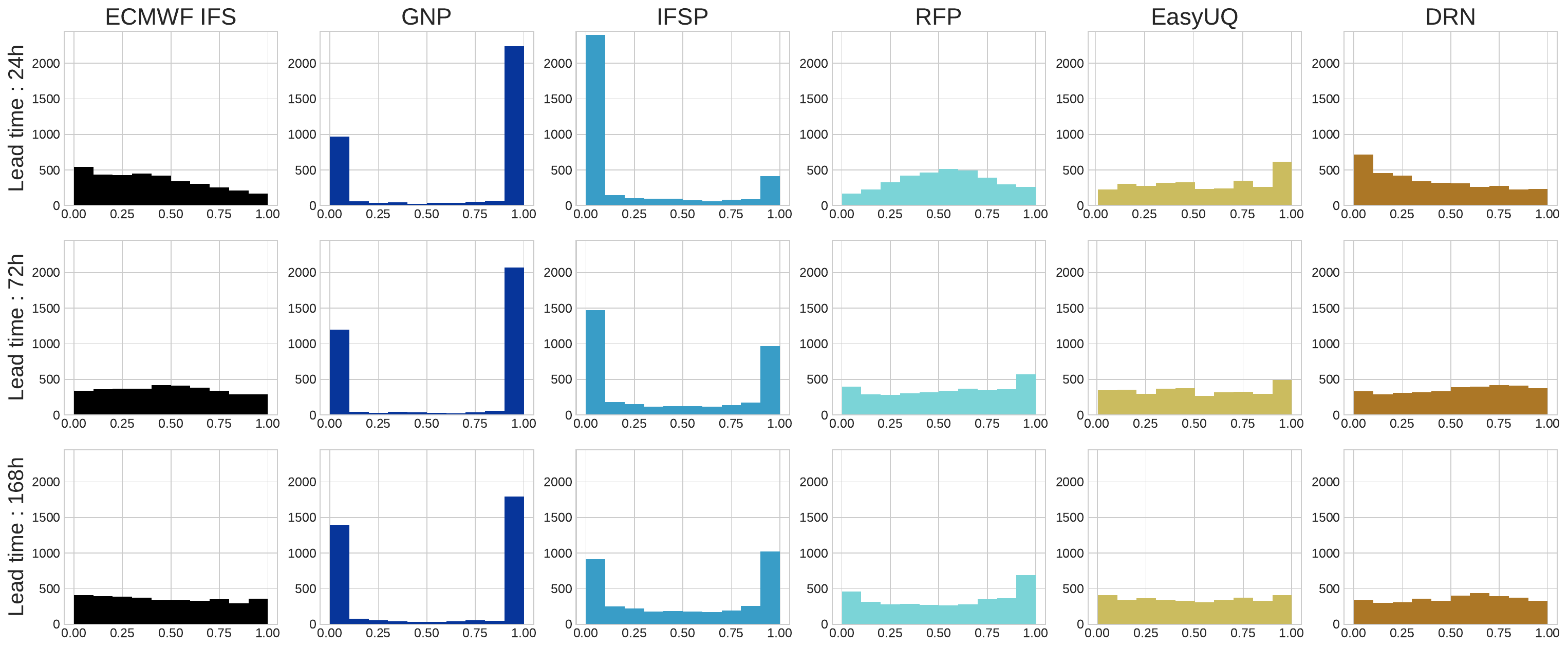}
\end{subfigure}
\caption{PIT histograms for all UQ methods and selected target variables. The results are aggregated over all test cases at $10$ randomly chosen grid points. The rows correspond to specific forecast lead times.}
\label{fig:pangu_pit}
\end{figure}

To investigate the calibration of the UQ methods, Figure \ref{fig:pangu_pit} shows PIT histograms for selected target variables and 10 randomly chosen grid points.
Most notable is the clear underdispersion of the GNP, IFSP and RFP approaches for the surface variables (U10 and T2M). 
For Z500, the GNP and IFSP forecasts are also underdispersed and show an additional bias, whereas the RFP forecasts are better calibrated.
The ECMWF ensemble forecasts are relatively well calibrated for most combinations of target variable and lead time, but tend to show minor underdispersion and biases.
The best calibration can be observed for the PH methods, apart from minor biases of DRN for Z500 forecasts at a lead time of 24 hours.

\begin{figure}
\centering
\begin{subfigure}[b]{0.49\textwidth}
    \caption{T2M}
   \includegraphics[width=1\linewidth]{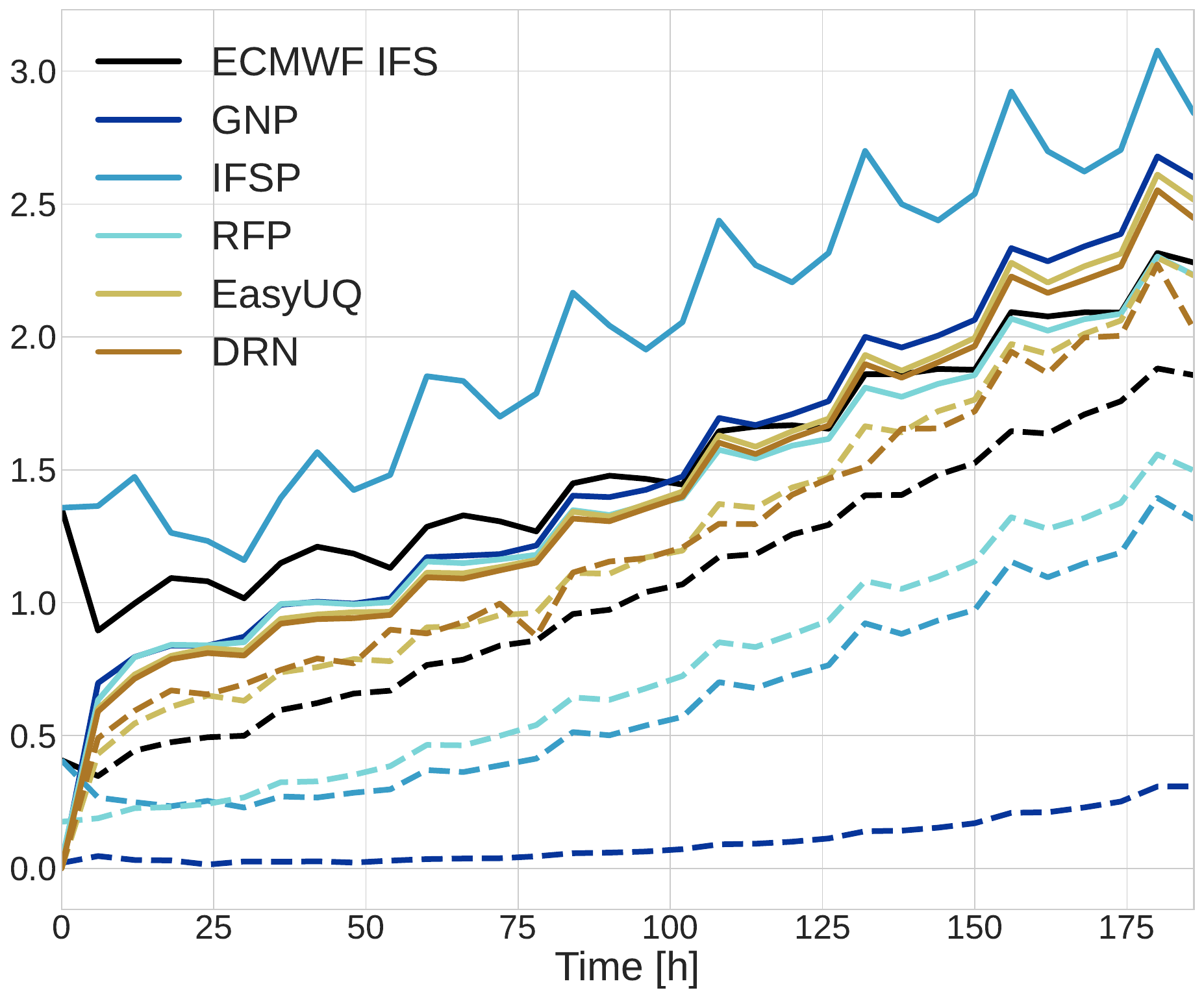}
\end{subfigure}
\begin{subfigure}[b]{0.49\textwidth}
    \caption{T850}
   \includegraphics[width=1\linewidth]{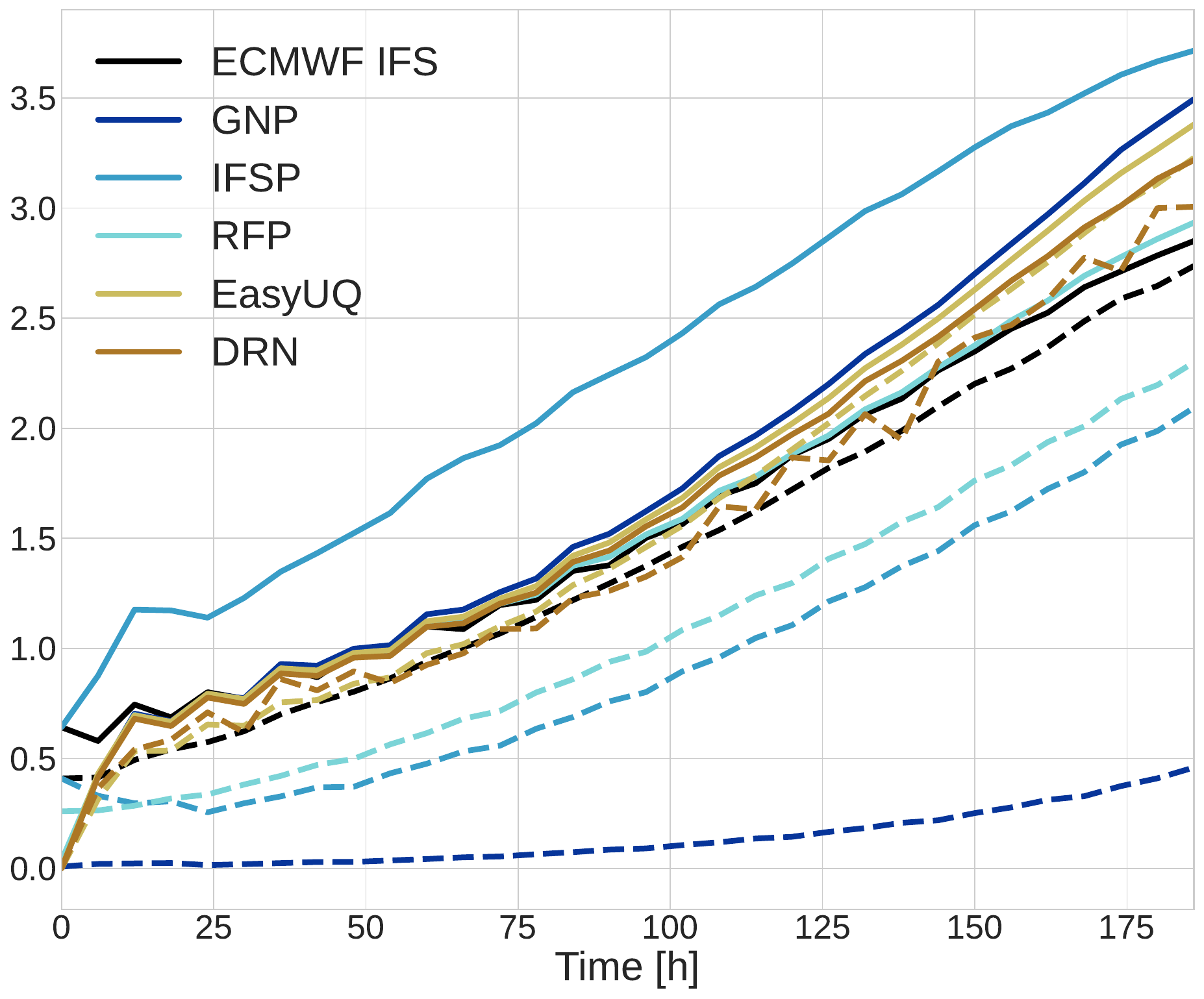}
\end{subfigure}
\caption{Spread-skill relationship between the RMSE of the ensemble mean and the ensemble spread of the different UQ methods, averaged over all grid points. The solid lines represents the RMSE, and the dotted line represents the standard deviations of the different methods, respectively.}
\label{fig:pangu_rmse_spread}
\end{figure}

A complementary perspective on calibration is provided by the spread-skill plots in Figure \ref{fig:pangu_rmse_spread}, which shows the relationship between the RMSE of the mean forecast and the standard deviation of the ensemble predictions of the different UQ methods for the two temperature variables.
For a well-calibrated ensemble forecast, the average ensemble spread should be roughly equal to the RMSE of the ensemble mean predictions for each lead time \citep{FortinEtAl2014}. As Figure \ref{fig:pangu_rmse_spread} indicates, this is not the case for most of the UQ methods considered here \footnote{In order to make the post-processing methods comparable, the predicted mean and standard deviation at each grid point were extracted from the forecasts to compute the RMSE and spread, respectively.}.
In particular for the IC methods, the standard deviation is notably lower than the RMSE, indicating a clear underestimation of the true forecast uncertainty by these approaches. It is noteworthy that the standard deviation of the IFSP forecasts actually decreases during the first 24\,h. This is somewhat surprising as the singular vectors at ECMWF represent the fastest growing perturbations over an optimization time window of 48 hours. Thus, a growth of the standard deviation would be expected. A similar behavior with initially slow perturbation growth was already documented in \citet{selz_2023} when Pangu-Weather was initialized with rescaled perturbations from the members of the ECMWF ensemble data assimilation. We attribute the different standard deviation growth between RFP and IFSP to the fact that the RFP method leads to perturbations which are larger in scale and magnitude than the IFSP perturbations (Fig.~\ref{fig:perturbation_example}) and thus grow faster initially.  
The two post-hoc methods show a similar behavior, with a slight underdispersion but in general good calibration, for short lead times even better than the ECMWF ensemble forecast. It should be kept in mind though that the ECMWF ensemble forecasts also contain the stochastically perturbed parametrization tendency scheme \citep{Buizza1999}, which leads to a better calibrated ensemble. 

\begin{figure}[p]
\centering
\begin{subfigure}[b]{\textwidth}
    \caption{U10}
   \includegraphics[width=1\linewidth]{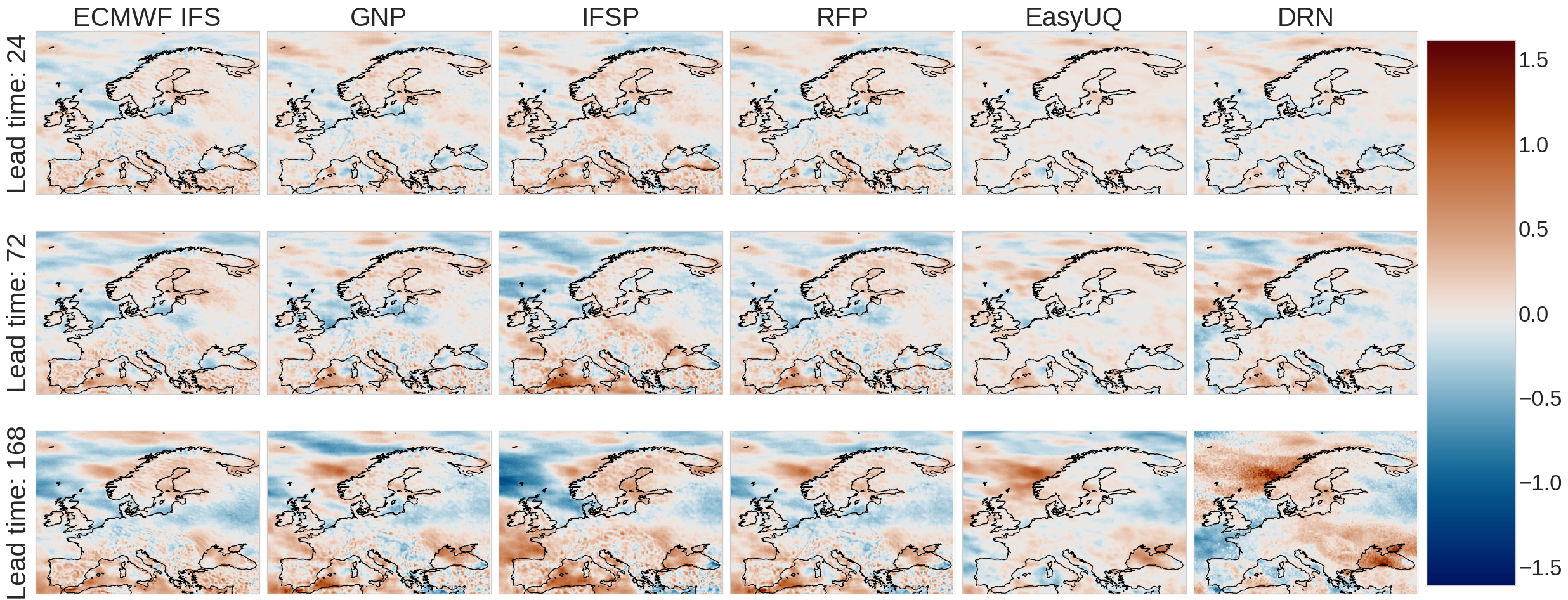}
\end{subfigure}
\begin{subfigure}[b]{\textwidth}
    \caption{T2M}
   \includegraphics[width=1\linewidth]{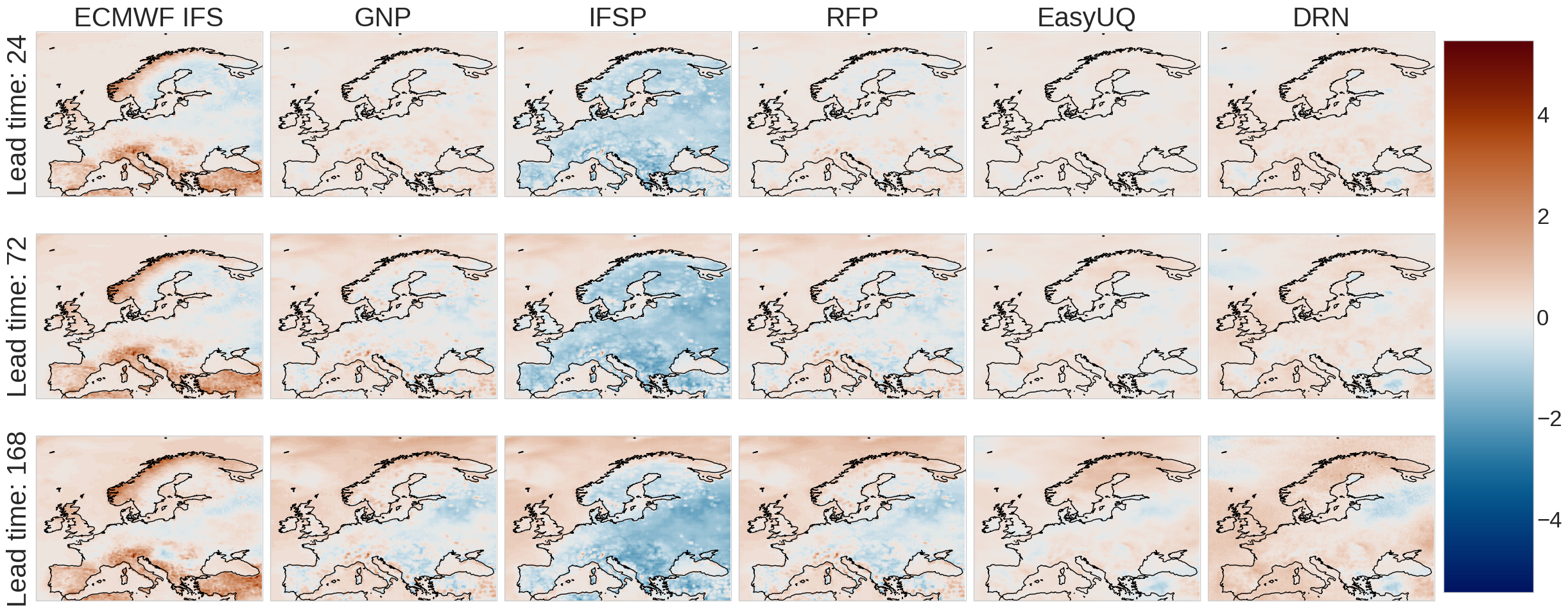}
\end{subfigure}
\begin{subfigure}[b]{\textwidth}
    \caption{Z500}
   \includegraphics[width=1\linewidth]{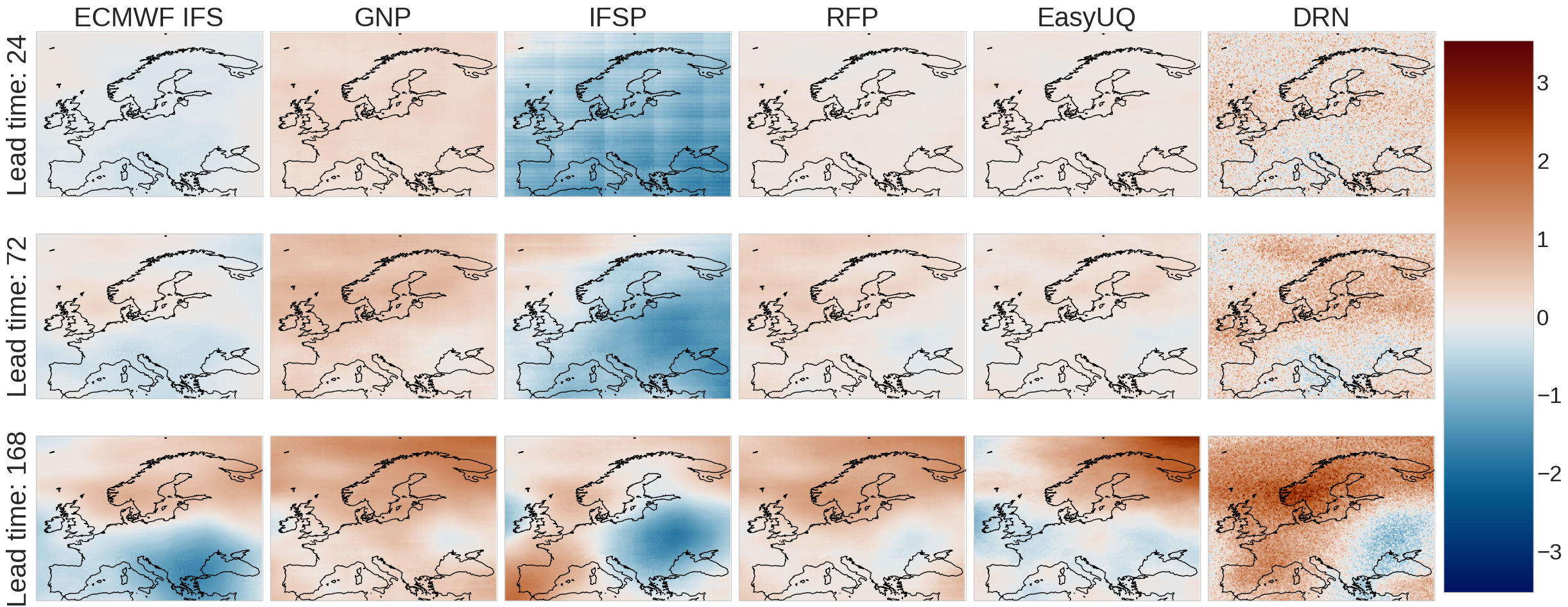}
\end{subfigure}
\caption{Bias of the mean forecast of different UQ methods for different lead times, averaged over the test period. }
\label{fig:pangu_bias_all}
\end{figure}

Figure \ref{fig:pangu_bias_all} shows the mean bias for forecasts of selected target variables for all UQ methods, where the bias is computed as the difference between the realizing observation and the mean forecast. A strong negative bias in the IFSP forecasts is apparent already at shorter lead times for T2M and, in particular, Z500. 
The likely cause of this bias, which, in turn, explains the observed bad performance of the IFSP approach, are systematic differences between the initial conditions of the operational ECMWF ensemble and the ERA5 reanalysis fields, which we considered as ground truth. 
Further, there are differences in the representation of topography in the operational version of the IFS model and ERA5 due to differences in the underlying grid spacing. 
One future pathway to addressing this, which has been suggested for example in \citet{WB2}, is to evaluate the NWP-based forecasts (including, potentially, the IFSP forecasts) against the operational analysis.
Although both post-processing methods operate separately on every grid point, only the DRN approach shows a strong granular pattern, while the bias pattern of the EasyUQ method appears notably more smooth and comparable to that of the RFP approach. The GNP method shows fairly small biases which are on a comparable level to those of the better-performing RFP and DRN approaches. 
While no substantial differences in the overall level of the bias among these methods can be observed, the DRN forecasts for Z500 at a lead time of 168 hours interestingly show the most pronounced biases among all compared methods.

\subsection{Comparison to post-processed IFS forecasts}\label{sec:IFSpp}

\begin{table}
	\centering
	\caption{Mean CRPS for the post-processing methods applied to IFS and Pangu-Weather predictions across the spatial domain for three different groups of lead times, with the best-performing method highlighted in bold. Note that the CRPS values for Z500 are scaled by a factor of $0.01$.}
	\label{tab:results_ifs_processed}
	\begin{tabular}{ c c|c c c| c c } 
		\toprule
		& Variable & IFS & IFS+EUQ & IFS+DRN & Pangu+EUQ & Pangu+DRN \\
		\midrule
		\multirow{5}{5em}{6h - 48h} & U10 & 0.54 & 0.53 & \textbf{0.51} & 0.53 & \textbf{0.51} \\
		&V10  & 0.54 & 0.53 & \textbf{0.51} & 0.53 & \textbf{0.51} \\
		&T2M  & 0.57 & 0.47 & 0.44 & 0.60  & \textbf{0.41} \\
		&T850 & 0.43 & 0.48 & 0.42 & 0.57  & \textbf{0.41} \\
		&Z500 & 0.33 &0.33 &\textbf{0.30}  & 0.36 & 0.32 \\
		\midrule
		
		\multirow{5}{5em}{48h - 120h} & U10 & \textbf{0.96} & 0.98 & \textbf{0.96} & 1.05 & 1.03 \\
		& V10 & \textbf{0.96} & 0.99 & \textbf{0.96}  & 1.05 & 1.03 \\
		& T2M & 0.75 &0.66 & \textbf{0.64}  & 0.69 & 0.67 \\
		& T850 & \textbf{0.75}& 0.78 & \textbf{0.75} & 0.82 & 0.79 \\
		& Z500 & \textbf{1.21} &1.27 & 1.22 & 1.35 & 1.29 \\
		\midrule
		
		\multirow{5}{5em}{$\geq$ 120h} & U10 & \textbf{1.54} & 1.57 & 1.55 & 1.70 & 1.68 \\
		& V10 & \textbf{1.58}& 1.61 & 1.59& 1.74 & 1.71 \\
		& T2M & 1.05& 1.00 & \textbf{0.98}& 1.13 & 1.10 \\
		& T850 &\textbf{1.33}&1.38 & 1.35 & 1.55 & 1.48 \\
		& Z500 & \textbf{2.91}&3.03 & 2.97 & 3.36 & 3.25 \\
		\bottomrule
	\end{tabular}
\end{table}

\begin{figure}
	\centering
	\includegraphics[width = \linewidth]{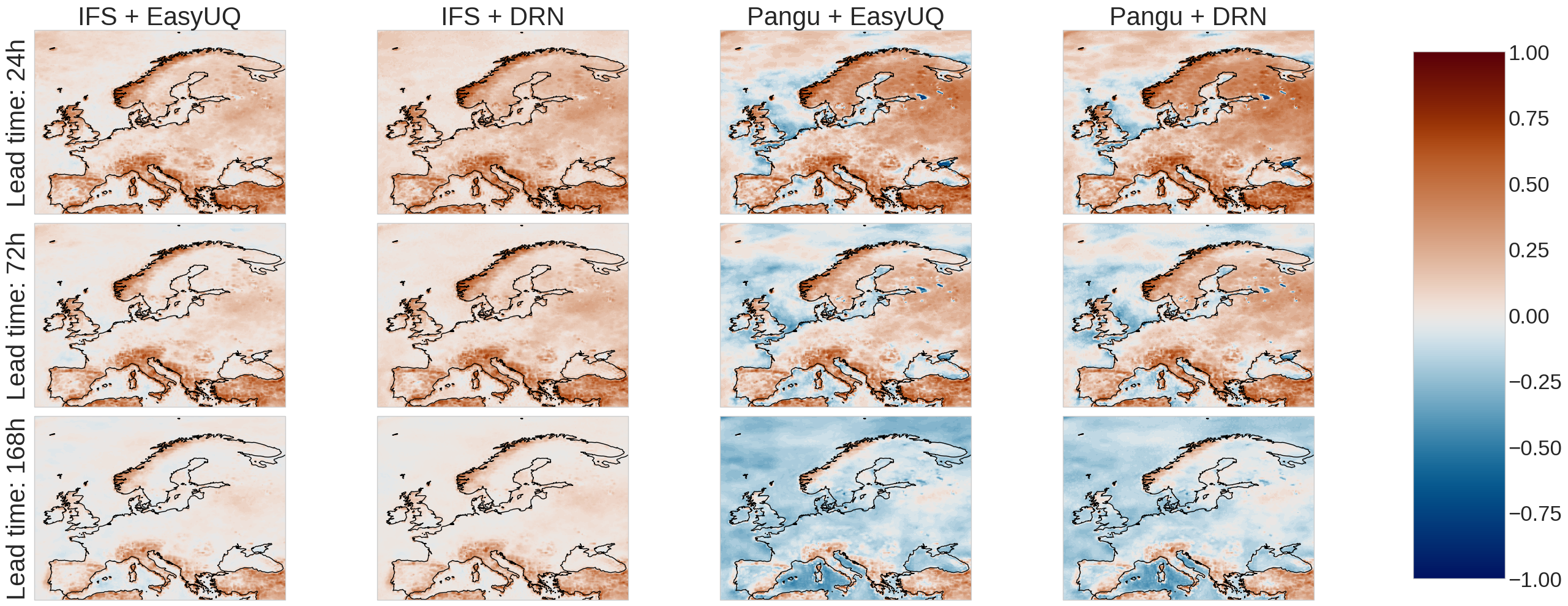}
	\caption{
		CRPSS of the different T2M post-processing methods applied to IFS and Pangu-Weather predictions over the spatial domain, using the ECMWF ensemble as a reference method. The rows correspond to specific forecasting lead times. Note that positive CRPSS values indicate an improvement over the reference in terms of the CRPS at the respective grid point.}
	\label{fig:crpss_ifs}
\end{figure}

Correcting systematic errors in NWP ensemble predictions was the original motivation for the development of post-processing methods such as DRN, and notable improvements in terms of the CRPS have been observed in numerous studies \citep{VannitsemEtAl2021}. 
As post-processing approaches have been widely adopted to improve physics-based NWP forecasts, post-processed IFS predictions thus constitute a natural benchmark for the UQ methods.
To compare post-processed data-driven and physics-based weather models, we applied the EasyUQ and DRN approach to the deterministic IFS forecast obtained from WeatherBench 2 \citep{WB2}, with the same specification and experimental setup as for the Pangu-Weather predictions. 
To enable a direct and fair comparison, we here utilize the deterministic IFS forecast only, and leave a comparison to post-processed ECMWF ensemble forecasts for future work.

Table \ref{tab:results_ifs_processed} summarizes the mean CRPS values for a direct comparison of the post-processed IFS and Pangu-Weather predictions. 
For shorter lead times of 6--48 hours, applying DRN to Pangu-Weather shows minimally better results than DRN applied to the IFS predictions. More notable differences (in favor of the post-processed IFS predictions) can be observed for EasyUQ. For lead times between 48 and 120 hours, IFS+DRN achieves the best performance and post-processed IFS forecasts show clear improvements over post-processed Pangu-Weather forecasts. That said, the post-processed IFS predictions only offer clear improvements over the raw ECMWF ensemble forecasts for T2M.
For lead times above 120 hours, the ECMWF ensemble forecasts perform best for most variables, with the exception of T2M where post-processing leads to improvements.

Figure \ref{fig:crpss_ifs} shows the CRPS skill score for the aforementioned post-processing methods for T2M. 
While post-processed Pangu-Weather forecasts obtain sizable improvements in high altitude areas and seem to perform better than post-processed IFS forecasts over land grid points for shorter lead times, their predictive performance over the sea is notably worse.
This comparison highlights that skillful probabilistic forecasts can be obtained by applying post-processing to both physics-based and data-driven weather models, with the best-performing combination varying across the meteorological variable, lead time and location.

\section{Discussion and conclusions}
\label{sec:conclusion}

To the best of our knowledge, our study is the first systematic comparison of different UQ methods to generate probabilistic weather forecasts from the deterministic data-driven weather model Pangu-Weather \citep{BiEtAl2023}.
The UQ approaches can be divided into initial condition-based methods, where an ensemble forecast is generated by initializing Pangu-Weather model runs from different sets of initial conditions, and post-hoc methods, which operate on deterministic Pangu-Weather forecasts and generate probabilistic forecasts from the deterministic model inputs, based on past forecasts and corresponding observations.
Overall, our results suggest that most of the UQ methods are able to provide probabilistic forecasts that are competitive with the operational (raw and post-processed) ECMWF ensemble forecast.

While the results differ substantially by variable and forecast lead time, the RFP, EasyUQ and DRN approaches perform generally similar to the operational ECMWF ensemble, while the GNP and IFSP approaches fail to achieve comparable forecast skill.
The most notable improvements over the ECMWF ensemble are achieved for 2-m temperature, where the use of the Pangu-Weather model in concert with UQ methods yields improvements in terms of the CRPS for lead times up to around 120 hours.
Generally, the PH methods (EasyUQ and, in particular, DRN) yield the best forecasts at shorter lead times, whereas the RFP approach yields better forecasts at longer lead times.
As discussed in Section \ref{sec:methods}\ref{sec:eval}, we use the ERA5 data as ground truth for evaluation throughout, and the model rankings might change if the operational analysis was used instead.

Our evaluation has been restricted to separately considering individual meteorological variables and grid points, and does not take into account spatial or inter-variable dependencies.
The IC approaches have the advantage that they generate realistic spatial forecast fields, as the input is perturbed over the whole spatial domain, whereas the PH methods generate a separate predictive distribution at every grid point. 
In particular, the RFP approach seems promising in that IC ensembles can be straightforwardly generated from past observation data with minimal tuning.
However, the use of the IC methods comes at the cost of having to run the deterministic data-driven weather model multiple times, which can be demanding in terms of the computing and disk space requirements, in particular for generating and storing global high-resolution ensemble forecasts.
By contrast, PH methods require a training dataset of past forecasts from the deterministic data-driven weather model and corresponding observations.
While they have the advantage of potentially correcting systematic errors such as biases of the underlying deterministic model, additional modeling steps are required to generate spatially coherent forecast fields.
A variety of two-step methods for multivariate post-processing is available, where in a first step, forecasts are post-processed separately at every grid point or lead time (using methods like, e.g., EasyUQ or DRN). 
In a second step, multivariate (e.g., spatial or temporal) dependencies are introduced by re-ordering samples from the univariate forecast distribution according to a dependence template via the use of copula functions.
Popular approaches include the use of empirical copulas based on the physics-based NWP ensemble models \citep[ensemble copula coupling, ECC;][]{ecc}, or based on past observations \citep[Schaake shuffle;][]{clark.2004}.
Comprehensive comparisons are for example available in \citet{LerchEtAl2020} and \citet{LakatosEtAl2023}.
In the context of post-processing data-driven weather model forecasts, the use of the ECC method comes with the benefit of obtaining a hybrid combination of a data-driven model producing the univariate forecasts at each grid point, and a physics-based ensemble model that provides information on the spatio-temporal dependencies.
However, ECC would still require physics-based NWP ensemble forecasts, in contrast to using Schaake shuffle to determine decencies from past observations.
Recently proposed multivariate post-processing methods based on generative ML \citep{ChenEtAl2022} further have the potential to better utilize various sources of input information and improve the multivariate probabilistic forecasts.

The main objective of our study was to provide a general proof of concept for how to generate probabilistic forecasts from deterministic data-driven weather models. 
It should be seen as a first step into the direction of probabilistic data-driven weather models, and our results provide several avenues for further generalization and analysis.
As a natural benchmark, we also applied the PH methods to the physics-based IFS forecast.
Our results indicate that at least for shorter lead times, the performance of post-processed IFS forecasts is in general quite similar to the post-processed data-driven forecast or sometimes even worse. These findings are in line with results of \citet{bremnes2023evaluation}, who compare post-processed Pangu-Weather and physics-based weather forecasts on a station dataset over Norway and find that the forecast quality tends to be very similar after post-processing. Comprehensive comparisons of post-processed data-driven and physics-based weather forecasts are an interesting starting point for future research, in particular also regarding the benefits of having an ensemble of NWP predictions available as input. Further, it would also be of interest to investigate whether post-processing methods could help to further improve the predictions of the IC-based UQ methods applied to deterministic data-driven weather forecasts we considered, for example by correcting some of the deficiencies observed for the GNP and IFSP approaches.

Thus far, our comparisons have been focused on selected target variables, on using the gridded ERA5 data as ground truth, and on the CRPS as main evaluation metric.
Operational weather services tend to evaluate their forecasts against  the model's own operational analysis, station observations \citep{WB2}, and comprehensive comparisons of the UQ methods constitute an interesting starting point for future research.
Over Europe, suitable station observation data has for example been collected within the EUPPBench benchmark dataset for post-processing \citep{DemaeyerEtAl2023}.
Another important open question regarding the potential and limitations of data-driven weather models is whether they can reliably predict extreme weather events.
Therefore, a targeted evaluation of the UQ methods in this regard, e.g., using proper weighted scoring rules \citep{LerchEtAl2017}, represents another important direction for future model comparisons.

The large data volumes and high dimensionality of global gridded predictions further poses a challenge regarding the scalability of ML-based post-processing methods such as DRN, for which it is an open question whether they will generalize well to global high-resolution forecasts.
This calls for the development of new spatial post-processing methods which operate on the spatial forecast fields directly and are able to leverage predictive information present in the spatial structures, as well as for the development of suitable evaluation metrics.
Over the past years, several approaches have been proposed, which utilize convolutional neural network or transformer architectures to enable probabilistic post-processing of spatial forecast fields \citep[e.g.,][]{GroenquistEtAl2021,AshkboosEtAl2022,ChapmanEtAl2022,PoET,HoratLerch2023}.
The recently introduced WeatherBench 2 dataset \citep{WB2} provides a useful framework for comparisons. In addition, future comparison should include inherently probabilistic data-driven models, such as GenCast \citep{GenCast},  NeuralGCM \citep{NeuralGCM}, or FuXi-ENS \citep{FuXiENS}. Although these methods already provide a data-driven ensemble forecast, their performance could potentially still be improved by applying additional post-processing for selected variables. 

As discussed above, the IC approaches are generally disadvantaged by their inability to account for sources of uncertainty beyond initial condition uncertainty.
One approach to address this might be to add scaled-down IC uncertainty information during the forward integration of the data-driven weather model, for example based on the use of perturbations determined from past analysis states.
Further, online bias correction or post-processing during the forward integration might help to alleviate systematic errors such as those observed for the IFSP approach and might constitute an interesting approach for combining the advantages of IC and PH methods.

\section*{Acknowledgments}

The research leading to these results has been done within the project ``Data-driven weather models: Towards improved uncertainty quantification, interpretability and efficiency'' funded by the Young Investigator Network at KIT. 
Christopher Bülte, Nina Horat and Sebastian Lerch gratefully acknowledges support by the Vector Stiftung through the Young Investigator Group ``Artificial Intelligence for Probabilistic Weather Forecasting''. The contribution of Julian Quinting was funded by the European Union (ERC, ASPIRE, 101077260).
We thank Delong Chen, Jieyu Chen, Charlotte Debus, Tilmann Gneiting, Christian Grams, Peter Knippertz, Linus Magnusson, and Jannik Wilhelm for helpful discussions. ECMWF and Deutscher Wetterdient are acknowledged for granting access to the operational ensemble forecast data. The authors acknowledge support by the state of Baden-Württemberg through bwHPC.

\bibliographystyle{myims2}
\bibliography{bibliography.bib}

\newpage
\appendix

\section{Results for FourCastNet}
\label{sec:fourcastnet}

Here, we present additional results for the previously introduced UQ methods utilizing the FourCastNet \citep{Pathak2022} model as the underlying data-driven weather model. For this purpose, FourCastNet version v0.0.0 was used, based on code accompanying the original publication\footnote{\url{https://github.com/NVlabs/FourCastNet}}.

\begin{table}[hb]
	\caption{Mean CRPS of all methods and variables across the spatial domain for three different groups of lead times, with the best-performing method highlighted in bold, respectively. The results shown here are analogous to those in Table \ref{tab:results_pangu}, but based on the FourCastNet model. Note that the CRPS values for Z500 are scaled by a factor of $0.01$.}
	\label{tab:results_fcn}
	\centering
	\begin{tabular}{ c c|c| c c c c c } 
		\toprule
		& Variable & ECMWF IFS  & GNP & IFSP &  RFP & EasyUQ & DRN \\
		\midrule
		\multirow{5}{5em}{Short time \\ 0h - 48h} & U10 & \textbf{0.54} & 0.67 & 0.68 & 0.64 & 0.62 & 0.60 \\
		& V10 & \textbf{0.54 }& 0.67 & 0.68 & 0.63 & 0.61 & 0.59 \\
		& T2M & 0.57 & 0.66 & 0.73 & 0.62 & 0.56 & \textbf{0.54} \\
		& T850 &\textbf{ 0.43} & 0.56 & 0.57 & 0.51 & 0.52 & 0.49 \\
		& Z500 &\textbf{ 0.33} & 0.74 & 0.71 & 0.78 & 0.68 & 0.72 \\
		\midrule
		
		\multirow{5}{5em}{Mid time \\ 48h - 120h} & U10 & \textbf{0.96} & 1.35 & 1.41 & 1.30 & 1.38 & 1.35 \\
		& V10 & \textbf{0.96 }& 1.37 & 1.43 & 1.32 & 1.39 & 1.37 \\
		& T2M & \textbf{0.75 }& 1.01 & 1.07 & 0.97 & 0.95 & 0.93 \\
		& T850 &\textbf{ 0.75} & 1.13 & 1.18 & 1.09 & 1.16 & 1.12 \\
		& Z500 &\textbf{ 1.21} & 2.44 & 2.50 & 2.35 & 2.42 & 2.40 \\
		\midrule
		
		\multirow{5}{5em}{Long time \\ 120h+} & U10 & \textbf{1.54} & 1.93 & 2.04 & 1.87 & 1.99 & 1.97 \\
		& V10 & \textbf{1.58 }& 1.99 & 2.11 & 1.93 & 2.05 & 2.01 \\
		& T2M & \textbf{1.05 }& 1.44 & 1.51 & 1.38 & 1.45 & 1.42 \\
		& T850 &\textbf{ 1.33} & 1.86 & 1.98 & 1.80 & 2.02 & 1.92 \\
		& Z500 &\textbf{ 2.91} & 4.44 & 4.76 & 4.31 & 4.80 & 4.66 \\
		\bottomrule
	\end{tabular}
\end{table}

\begin{figure}[p]
    \centering
    \includegraphics[width = \textwidth]{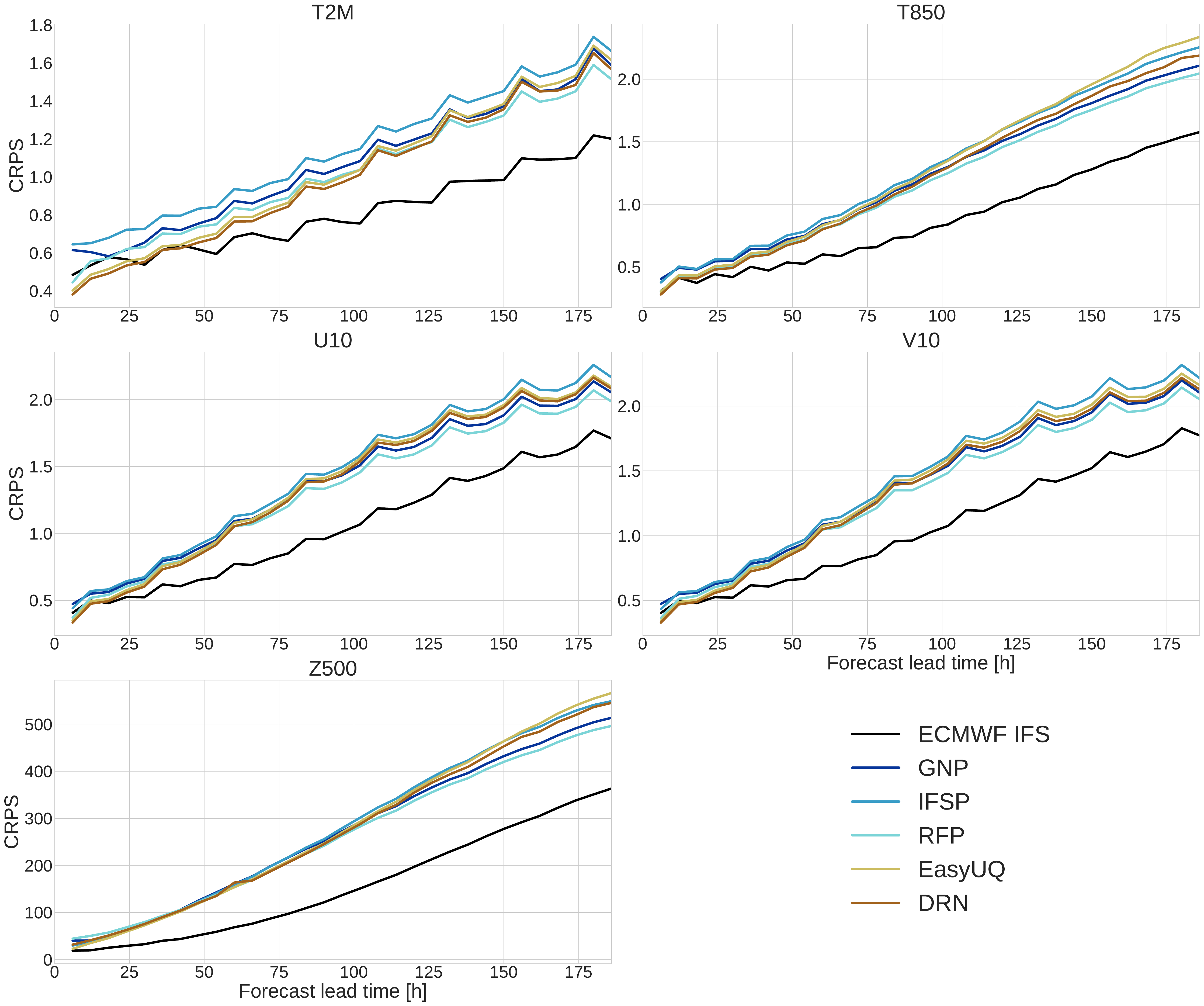}
    \caption{Mean CRPS as a function of the forecast lead time for the different UQ methods, aggregated over all locations. The results shown here are analogous to those in Figure \ref{fig:crps_per_time_pangu}, but based on the FourCastNet model.}
    \label{fig:crps_per_time_fcn}
\end{figure}

\autoref{tab:results_fcn} shows the mean CRPS results over all test samples for different groups of lead times, and Figure \ref{fig:crps_per_time_fcn} shows the mean CRPS as a function of the lead time. 
Overall, we observe qualitatively similar results to the probabilistic forecasts based on the Pangu-Weather model, but the forecast quality is notably worse.
This is likely due to the worse forecast performance of the underlying FourCastNet model compared to Pangu-Weather.
While analogous rankings between IC and PH approaches can be observed, forecasts of T2M and lead times below around 50 hours are the only case among all considered methods and target variables, where any of the UQ methods can achieve any improvements over the operational ECMWF ensemble.

\begin{figure}[p]
\centering
\begin{subfigure}[b]{\textwidth}
    \caption{U10}
   \includegraphics[width=1\linewidth]{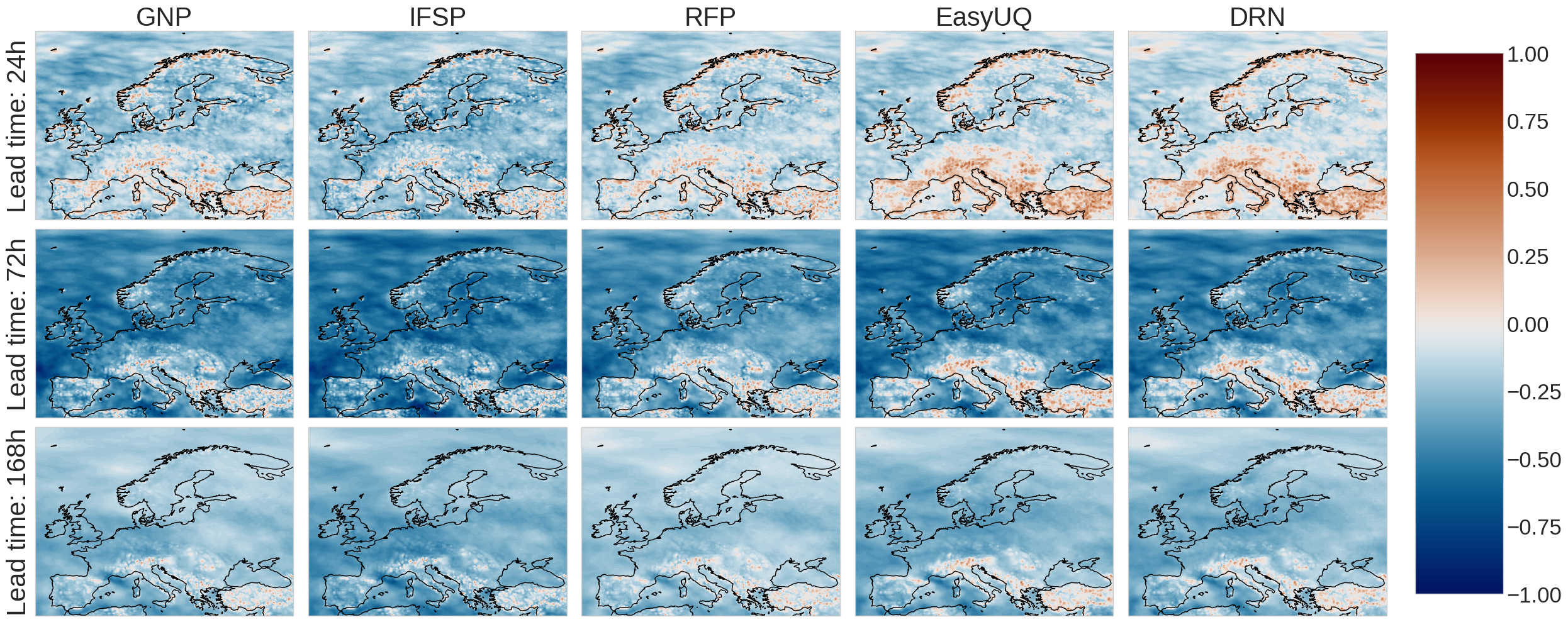}
\end{subfigure}
\begin{subfigure}[b]{\textwidth}
    \caption{T2M}
   \includegraphics[width=1\linewidth]{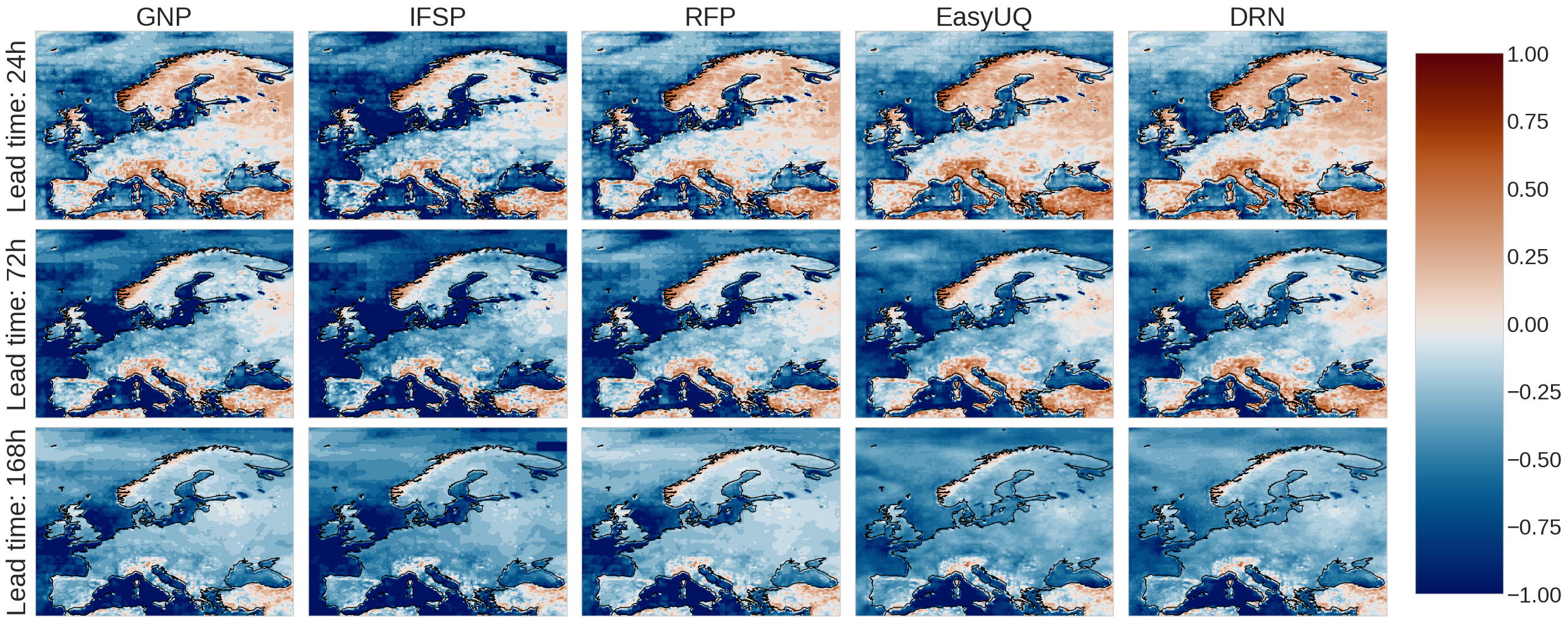}
\end{subfigure}
\begin{subfigure}[b]{\textwidth}
    \caption{Z500}
   \includegraphics[width=1\linewidth]{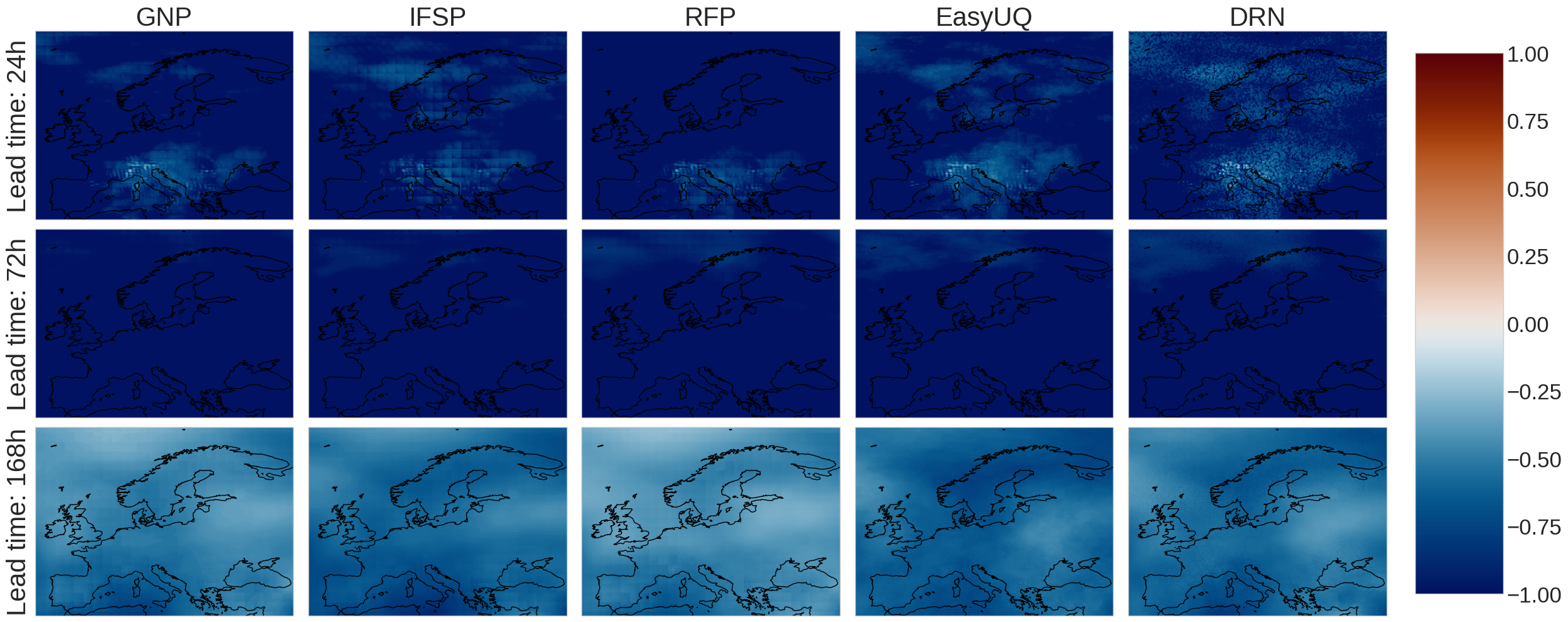}
\end{subfigure}
\caption{CRPSS of the different UQ methods over the spatial domain, using the ECMWF ensemble as a reference method. The rows correspond to specific forecasting lead times. Note that positive CRPSS values indicate an improvement over the reference in terms of the CRPS at the respective grid point. The results shown here are analogous to those in Figure \ref{fig:pangu_crpss}, but based on the FourCastNet model.}
\label{fig:fcn_crpss}
\end{figure}

\autoref{fig:fcn_crpss} shows the CRPSS of the different methods over the spatial domain for selected target variables and lead times.
Compared to the corresponding results for the Pangu-Weather model, the results are notably worse everywhere, but some improvements over the ECMWF ensemble can be observed over the land grid points, in particular at higher altitudes and for the PH methods.

\end{document}